\documentclass[a4paper,fleqn]{cas-sc}

\usepackage[T1]{fontenc}
\usepackage[polutonikogreek,italian,english]{babel}
\usepackage[page,toc,titletoc,title]{appendix}
\usepackage[numbers,sort&compress]{natbib}
\usepackage{graphicx}
\usepackage{graphics}
\usepackage{braket}
\usepackage{bbold}
\usepackage{amssymb}
\usepackage{amsmath}
\usepackage{nicefrac}
\usepackage{dcolumn}
\usepackage{bm}
\usepackage{slashed}
\usepackage{datetime}
\usepackage{mciteplus}
\usepackage{multirow}
\usepackage{bbold}
\usepackage{color}
\usepackage{xcolor}
\usepackage{float}
\usepackage[utf8]{inputenc}
\usepackage[normalem]{ulem}
\usepackage{acro}

\DeclareAcronym{AI}{
  short=AI,
  long=Artificial Intelligence,
}
\DeclareAcronym{BSM}{
  short=BSM,
  long=Beyond-the-Standard-Model,
}
\DeclareAcronym{QCD}{
  short=QCD,
  long=Quantum ChromoDynamics,
}
\DeclareAcronym{SM}{
  short=SM,
  long=Standard Model,
}
\DeclareAcronym{CP}{
  short=CP,
  long=Charge-Parity,
}
\DeclareAcronym{SLAC}{
  short=SLAC,
  long=Stanford Linear Accelerator Center,
}
\DeclareAcronym{BNL}{
  short=BNL,
  long=Brookhaven National Laboratory,
}
\DeclareAcronym{FCNCs}{
  short=FCNCs,
  long=Flavor-Changing Neutral Currents,
}
\DeclareAcronym{GIM}{
  short=GIM,
  long=Glashow--Iliopoulos--Maiani,
}
\DeclareAcronym{CEM}{
  short=CEM,
  long= Color Evaporation Model,
}
\DeclareAcronym{CSM}{
  short=CSM,
  long= Color Singlet Mechanism,
}
\DeclareAcronym{CO}{
  short=CO,
  long=Color Octet,
}
\DeclareAcronym{NRQCD}{
  short=NRQCD,
  long=NonRelativistic QCD,
}
\DeclareAcronym{LDME}{
  short=LDME,
  long=Long-Distance Matrix Element,
}
\DeclareAcronym{LO}{
  short=LO,
  long=Leading Order,
}
\DeclareAcronym{NLO}{
  short=NLO,
  long=Next-to-Leading Order,
}
\DeclareAcronym{NNLO}{
  short=NNLO,
  long=Next-to-NLO,
}
\DeclareAcronym{MHOUs}{
  short=MHOUs,
  long=Missing Higher-Order Uncertainties,
}
\DeclareAcronym{DIS}{
  short=DIS,
  long=Deep Inelastic Scattering,
}
\DeclareAcronym{DGLAP}{
  short=DGLAP,
  long=Dokshitzer--Gribov--Lipatov--Altarelli--Parisi,
}
\DeclareAcronym{PDFs}{
  short=PDFs,
  long=Parton Distribution Functions,
}
\DeclareAcronym{FFs}{
  short=FFs,
  long=Fragmentation Functions,
}
\DeclareAcronym{MPIs}{
  short=MPIs,
  long=Multi-Parton Interactions,
}
\DeclareAcronym{DPS}{
  short=DPS,
  long=Double-Parton Scattering,
}
\DeclareAcronym{SCET}{
  short=SCET,
  long=Soft and Collinear Effective Theory,
}
\DeclareAcronym{TM}{
  short=TM,
  long=Transverse-Momentum,
}
\DeclareAcronym{TMD}{
  short=TMD,
  long=Transverse-Momentum-Dependent,
}
\DeclareAcronym{FFNS}{
  short=FFNS,
  long=Fixed-Flavor Number Scheme,
}
\DeclareAcronym{VFNS}{
  short=VFNS,
  long=Variable-Flavor Number Scheme,
}
\DeclareAcronym{GM-VFNS}{
  short=GM-VFNS,
  long=General-Mass Variable-Flavor Number Scheme,
}
\DeclareAcronym{ABF}{
  short=ABF,
  long=Altarelli--Ball--Forte,
}
\DeclareAcronym{BFKL}{
  short=BFKL,
  long=Balitsky--Fadin--Kuraev--Lipatov,
}
\DeclareAcronym{LL}{
  short=LL,
  long=Leading Logarithmic,
}
\DeclareAcronym{NLL}{
  short=NLL,
  long=Next-to-Leading Logarithmic,
}
\DeclareAcronym{NNLL}{
  short=NNLL,
  long=Next-to-NLL,
}
\DeclareAcronym{LVM}{
  short=LVM,
  long=Light Vector Meson,
}
\DeclareAcronym{UGD}{
  short=UGD,
  long=Unintegrated Gluon Distribution,
}
\DeclareAcronym{LHC}{
  short=LHC,
  long=Large Hadron Collider,
}
\DeclareAcronym{EIC}{
  short=EIC,
  long=Electron-Ion Collider,
}
\DeclareAcronym{HFAG}{
  short=HFAG,
  long=Heavy Flavor Averaging Group,
}
\DeclareAcronym{SCA}{
  short=SCA,
  long=Small-Cone Algorithm
}
\DeclareAcronym{BLM}{
  short=BLM,
  long=Brodsky--Lepage--Mackenzie,
}
\DeclareAcronym{SNAJ}{
  short=SNAJ,
  long=Suzuki--Nejad--Amiri--Ji,
}

\newcommand{\deffont}[1]{\begin{otherlanguage*}{polutonikogreek}#1\end{otherlanguage*}}

\def\tsc#1{\csdef{#1}{\textsc{\lowercase{#1}}\xspace}}
\tsc{WGM}
\tsc{QE}

\newcommand{\drv}{{\rm d}}

\newcommand{\LQCD}{\Lambda_{\rm QCD}}
\newcommand{\MSb}{\overline{\rm MS}}

\newcommand{\NLO}{{\rm NLO}}

\newcommand{\LL}{{\rm LL/LO}}
\newcommand{\NLL}{{\rm NLL/NLO}}
\newcommand{\NLLp}{{\rm NLL/NLO^+}}
\newcommand{\NLLpp}{{\rm NLL/NLO^{(+)}}}
\newcommand{\HENLOp}{{\rm HE}\mbox{-}{\rm NLO^+}}

\newcommand{\CnLL}{{\cal C}_n^\LL}

\newcommand{\CnNLLp}{{\cal C}_n^\NLLp}

\newcommand{\CnHENLOp}{{\cal C}_n^{{\rm HE}\text{-}{\rm NLO}^+}}
\newcommand{\DY}{\Delta Y}

\newcommand{\F}{{\cal F}}

\newcommand{\HQ}{{\cal H}_Q}
\newcommand{\Hc}{{\cal H}_c}
\newcommand{\Hb}{{\cal H}_b}
\newcommand{\Jpsi}{J/\psi}
\newcommand{\Yps}{\Upsilon}

\newcommand{\XQq}{ X_{Qq\bar{Q}\bar{q}}}
\newcommand{\Xcq}{ X_{cq\bar{c}\bar{q}}}
\newcommand{\Xbq}{ X_{bq\bar{b}\bar{q}}}

\newcommand{\Xcu}{ X_{cu\bar{c}\bar{u}}}
\newcommand{\Xcs}{ X_{cs\bar{c}\bar{s}}}
\newcommand{\Xbu}{ X_{bu\bar{b}\bar{u}}}
\newcommand{\Xbs}{ X_{bs\bar{b}\bar{s}}}

\newcommand{{\Jethad}}{\textsc{Jethad}}
\newcommand{{\Hell}}{\textsc{Hell}}
\newcommand{{\RadISH}}{\textsc{RadISH}}

\newcommand{\tref}[1]{~\ref{#1}}

\begin{document}
\let\WriteBookmarks\relax
\def\floatpagepagefraction{1}
\def\textpagefraction{.001}

\shorttitle{Exotic tetraquarks at the HL-LHC with {\Jethad}: A high-energy viewpoint}    

\shortauthors{Celiberto, Francesco Giovanni}  

\title []{\Huge Exotic tetraquarks at the HL-LHC with {\Jethad}: \\ A high-energy viewpoint}  

\author[1]{Francesco Giovanni Celiberto}[orcid=0000-0003-3299-2203]

\cormark[1]


\ead{francesco.celiberto@uah.es}


\affiliation[1]{organization={Universidad de Alcal\'a (UAH), Departamento de F\'isica y Matem\'aticas},
            addressline={Campus Universitario}, 
            city={Alcal\'a de Henares},
            postcode={E-28805}, 
            state={Madrid},
            country={Spain}}




\begin{abstract}
We review the semi-inclusive hadroproduction of a neutral hidden-flavor tetraquark with light and heavy quark flavor at the HL-LHC, accompanied by another heavy hadron or a light-flavored jet.
We make use of the novel {\tt TQHL1.0} determinations of leading-twist fragmentation functions to describe the formation mechanism of a tetraquark state within the next-to-leading order perturbative QCD.
This framework builds on the basis of a spin-physics inspired model, taken as a proxy for the lowest-scale input of the constituent heavy-quark fragmentation channel. 
Then, all parton-to-tetraquark fragmentation functions are consistently obtained via the above-threshold DGLAP evolution in a variable-flavor number scheme.
We provide predictions for a series of differential distributions calculated by the hands of the {\Jethad} method well-adapted to $\NLLp$ hybrid-factorization studies, where the resummation of next-to-leading energy logarithms and beyond is included in the collinear picture.
We provide corroborating evidence that high-energy observables sensitive to semi-inclusive tetraquark emissions at the HL-LHC exhibit a fair stability under radiative corrections as well as MHOU studies.
Our analysis constitutes a prime contact point between QCD resummations and the exotic matter.
\end{abstract}




\maketitle

\newcounter{appcnt}


\tableofcontents
\clearpage

\setlength{\parskip}{3pt}%

\section{Hors d'{\oe}uvre}
\label{sec:introduction}

The study of heavy-quark flavored objects in high-energy hadron collisions is crucial for shedding light on the core nature of fundamental interactions. 
These processes serve as gold-plated channels for probing the underlying dynamics of particle physics. 
Heavy quarks, due to their large masses, act as sentinels for potential imprints of New Physics, as they could interact with \ac{BSM} particles. 
On the other hand, their masses fall within a regime where perturbative \ac{QCD} calculations are feasible, thus enabling precision studies of strong interactions.

QCD, the theory of the strong force, is a cornerstone of the \ac{SM} of particle physics. It is based on the non-Abelian $SU(N_c)$ gauge group, with $N_c=3$ representing the number of colors~\cite{Gell-Mann:1962yej,Gell-Mann:1964ewy,Zweig:1964jf,Fritzsch:1973pi}.
Quarks, which come in six flavors along with their antiquarks, constitute the building blocks of hadrons. In the QCD framework, quarks are described by fermionic fields belonging to the fundamental triplet representation of $SU(3)$. 
Gluons, the force carriers of the strong interaction, are massless spin-1 bosons that mediate interactions between quarks. 
In the QCD Lagrangian, gluons are represented by bosonic fields belonging to the adjoint octet representation of $SU(3)$.

While QCD stands as one of the fundamental pillars of the SM, it also serves as a fertile ground for probing BSM physics. Several potential portals to BSM extensions within the realm of QCD have been proposed.
Among others, they include: axions, which were originally postulated to address the strong \ac{CP} problem~\cite{Peccei:1977hh,Peccei:1977ur,Peccei:2006as,Duffy:2009ig}, non-Abelian dark gauge forces~\cite{Forestell:2017wov,Huang:2020crf}, quarkyonic matter~\cite{McLerran:2007qj,Hidaka:2008yy,McLerran:2018hbz}, and higher-dimensional QCD operators incorporated into the Lagrangian~\cite{Buchmuller:1985jz,Witten:1979kh,Dudek:2010wm,Afonin:2019unu}. 
This wealthy of possibilities provide us with rich opportunities for exploring the frontiers of particle physics and searching for long-awaited signs of physics beyond the SM.

Within the QCD domain, a key role is played by hadrons whose lowest Fock state contains two or more heavy quarks.
Mesons composed of a heavy quark ($Q$) and its antiquark ($\bar Q$) are known as (heavy) \emph{quarkonia}.
The origins of quarkonium studies trace back to the so-called ``Quarkonium November Revolution''.
Indeed, in November 1974, a groundbreaking discovery was made: a new vector meson, with a mass around 3.1 GeV and carrying photon quantum numbers, was observed independently by two research groups. This meson was named $\Jpsi$, after the two institutions where it was discovered: the Stanford Linear Accelerator Center (SLAC) led by B. Richter~\cite{SLAC-SP-017:1974ind}, and the Brookhaven National Laboratory (BNL) led by S. Ting~\cite{E598:1974sol}. 
Shortly after these announcements, the discovery of the $\Jpsi$ was confirmed by the Frascati ADONE experiment, under the direction of G. Bellettini~\cite{Bacci:1974za}. 
This landmark heralded the beginning of quarkonium studies and provided crucial insights into the nature of the strong force and the underlying quark structure of hadrons.

Though quarkonium mesons fall into the class of \emph{ordinary} hadrons, QCD color neutrality allows for the formation of bound states with more intricate valence-parton configurations, leading to the emergence of \emph{exotic} hadrons. 
These exotic particles have quantum numbers that cannot be explained by conventional quark-antiquark or three-quark configurations. 
Instead, they are composed of unique combinations of quarks, antiquarks, and gluons. Understanding the inner structure of these exotic hadrons has been the focus of intense research in the field of exotic spectroscopy.
Exotic hadrons can be classified into two main categories: those composed of active gluons, such as quark-gluon hybrids~\cite{Kou:2005gt,Braaten:2013boa,Berwein:2015vca} (see also Refs.~\cite{Szczepaniak:2015eza,Szczepaniak:2015hya,Guo:2015umn,Swanson:2015bsa,Guo:2019twa}) and glueballs~\cite{Close:1991pf,Close:1997qda,Close:1998zz,Minkowski:1998mf,Close:2000yg,Mathieu:2008me,Hsiao:2013dta,D0:2020tig,Csorgo:2019ewn}, and those composed of multiple quarks, such as tetraquarks and pentaquarks~\cite{Gell-Mann:1964ewy,Jaffe:1976ig,Jaffe:1976ih,Ader:1981db}.

The observation of the first exotic hadron, the $X(3872)$, occurred in 2003 at the Belle experiment~\cite{Belle:2003nnu}, and subsequent experiments confirmed its existence. 
Its discovery marked the beginning of the so-called ``Second Quarkonium Revolution'', or ``Exotic Revolution''.
The $X(3872)$ is a hidden-charm particle, believed to be composed of charm and anticharm quarks~\cite{Chen:2016qju,Liu:2019zoy,Esposito:2020ywk}.
The first exotic state with open-charm flavor, the $X(2900)$ particle, was observed for the first time in 2021 at LHCb~\cite{LHCb:2020bls}.
Although the $X(3872)$ has quantum numbers that are not exotic, its decay properties violate isospin conservation, suggesting that its inner structure may involve more complex dynamics beyond traditional quarkonium states. 
Various theoretical models have been proposed to describe the $X(3872)$, including a loosely-bound meson molecule~\cite{Tornqvist:1993ng,Braaten:2003he,Braaten:2010mg,Braaten:2020iye,Guo:2013sya,Guo:2013xga,Cleven:2015era,Fleming:2021wmk,Dai:2023mxm,Fleming:2007rp,Fleming:2008yn,Fleming:2011xa,Mehen:2015efa,Mutuk:2022ckn,Wang:2020dgr,Wang:2013daa,Xin:2021wcr,Wang:2020cme,Wang:2014gwa}, a compact diquark system~\cite{Maiani:2004vq,tHooft:2008rus,Maiani:2013nmn,Maiani:2014aja,Maiani:2017kyi,Mutuk:2021hmi,Mutuk:2021epz,Mutuk:2022zgn,Mutuk:2022nkw,Wang:2023sii,Wang:2019tlw,Wang:2013vex,Wang:2013llv,Wang:2013exa}, or a hadroquarkonium configuration consisting of a quarkonium core and an orbiting light meson~\cite{Dubynskiy:2008mq,Dubynskiy:2008di,Li:2013ssa,Voloshin:2013dpa,Guo:2017jvc,Ferretti:2018ojb,Ferretti:2018tco,Ferretti:2020ewe}.

Quite recently, a high-energy description of the single hadroproduction of fully charmed tetraquarks was proposed~\cite{Maciula:2020wri,Cisek:2022uqx}.
The production rates of the $X(3872)$ at large transverse momenta, as measured by experiments at the LHC, provide valuable insights into its production mechanisms. 
These measurements can help to constrain theoretical models and favor production mechanisms inherent in high-energy QCD, such as the fragmentation of a single parton into the observed particle. 

In this review we address the associated hadroproduction, at LHC and its high-luminosity upgrade (HL-LHC), of a forward heavy-light tetraquark and a backward singly heavy-flavored hadron or a light jet.
The two outgoing particles possesses high transverse momenta and a large mutual separation in rapidity. 
On the one side, the presence of moderate parton longitudinal-momentum fractions allows for a reliable description using collinear \ac{PDFs}.
On the other side, large rapidity intervals in the final state lead to significant exchanges of transverse momenta in the $t$-channel.

Therefore, a high-energy factorization treatment, accounting for energy logarithms due to $t$-channel gluon emissions, becomes necessary.
With the aim of providing an accurate high-energy QCD description of our reactions in these kinematic ranges, we rely upon the {\Jethad} method~\cite{Celiberto:2020wpk,Celiberto:2022rfj,Celiberto:2023fzz} to implement the $\NLLp$ hybrid-factorization formalism~\cite{Colferai:2010wu,Celiberto:2020wpk,Celiberto:2020tmb,Bolognino:2021mrc,Celiberto:2022rfj,Celiberto:2022dyf,Celiberto:2023fzz}, in which the resummation of next-to-leading energy logarithms and beyond is consistently included in the collinear picture.

Then, the last required ingredient is the choice of a reliable mechanism depicting the formation of our heavy-light tetraquark states in the kinematic regimes of interest.
From a collinear-factorization viewpoint, the transverse momenta at which final-state particles are tagged make the use of a \ac{VFNS} approach~\cite{Mele:1990cw,Cacciari:1993mq} valid.

To this extent, we will make use of a novel set of collinear \ac{FFs}, named {\tt TQHL1.0}, suited to  the VFNS fragmentation of $\XQq$ tetraquarks~\cite{Celiberto:2023rzw}.
They were obtained by performing a next-to-leading \ac{DGLAP} evolution of an initial-scale input for the constituent heavy-quark fragmentation channel, obtained within the spin-physics inspired \ac{SNAJ} model~\cite{Suzuki:1985up,Nejad:2021mmp,Suzuki:1977km,Amiri:1986zv}.

For the sake of completeness, we mention that other hybrid-factorization formalisms, closely related to our $\NLLp$ framework and tailored for single forward detections, exist. 
These include the approaches proposed in Refs.~\cite{Deak:2009xt,vanHameren:2015uia,Deak:2018obv,VanHaevermaet:2020rro,vanHameren:2022mtk,Giachino:2023loc,Guiot:2024oja}, which offer valuable insights and can complement our approach in understanding high-energy QCD processes.

On the other side, analyses on small-$x$ resummed inclusive or differential distributions for Higgs and heavy-flavor hadroproduction have been conducted using the {\Hell} method~\cite{Bonvini:2018ixe,Silvetti:2022hyc}. The {\Hell} method relies on the \ac{ABF} approach~\cite{Ball:1995vc,Ball:1997vf,Altarelli:2001ji,Altarelli:2003hk,Altarelli:2005ni,Altarelli:2008aj,White:2006yh}, which combines collinear factorization with small-$x$ resummation, along with high-energy factorization theorems~\cite{Catani:1990xk,Catani:1990eg,Collins:1991ty,Catani:1993ww,Catani:1993rn,Catani:1994sq,Ball:2007ra,Caola:2010kv}. These analyses offer complementary perspectives and contribute to a more comprehensive understanding of high-energy QCD phenomena.

The outline of this review reads as follows. 
In Section~\ref{sec:exotic_tetraquarks} we introduce the $\NLLpp$ hybrid collinear and high-energy factorization, whereas in we Section~\ref{sec:HF_fragmentation} we provide technical details on our strategy to describe tetraquark collinear fragmentation.
Results and conclusions are presented in Sections~\ref{sec:phenomenology} and~\ref{sec:conclusions}, respectively.

\section{Theoretical setup}
\label{sec:exotic_tetraquarks}

This Section is for a digression on recent phenomenological progresses of high-energy resummation in QCD (see Section~\ref{ssec:HE_QCD}).
Then, it gives a formal description of the observables matter of our analysis as cast within the hybrid factorization (see Section~\ref{ssec:hybrid_factorization}).

\subsection{High-energy QCD phenomenology: An incomplete summary}
\label{ssec:HE_QCD}

Providing accurate predictions for high-energy observables relies upon the ability to disentangle long-distance from short-distance dynamics in hadron scatterings. 
This permits to factorize nonperturbative dynamics from perturbative calculations via the well-established collinear framework~\cite{Collins:1989gx,Sterman:1995fz}.

However, specific kinematic regions pose challenges due to the emergence of large logarithms. 
These logarithmic corrections grows with the order of the perturbative expansion, thus offsetting the smallness of the QCD running coupling and hampering the convergence of perturbative series. 
In such scenarios, the standard collinear factorization must be enhanced \emph{via} the incorporation of all-order resummations.

In the semi-hard regime of QCD~\cite{Gribov:1983ivg} (see Refs.~\cite{Celiberto:2017ius,Bolognino:2021bjd,Celiberto:2022qbh} for advances in phenomenology), characterized by a stringent energy scale hierarchy $\sqrt{s} \gg {Q} \gg \LQCD$, large logarithms of the form $\ln s/Q^2$ enter perturbative series with a power increasing with the order, thus calling for a resummation treatment. 

The \ac{BFKL} formalism~\cite{Fadin:1975cb,Kuraev:1976ge,Kuraev:1977fs,Balitsky:1978ic} proves to be the most adequate tool for high-energy resummation.
More in particular, BFKL allows us to resum all  contributions proportional to $(\alpha_s \ln s )^n$, the so-called \ac{LL} approximation, as well as the ones going with to $\alpha_s(\alpha_s \ln s)^n$, namely the next-to-leading logarithmic \ac{NLL} approximation.

According to BFKL, a given scattering amplitude factorizes as a convolution between a universal Green's function and two singly-off-shell, transverse-momentum dependent emission functions (also known as impact factors) describing the emission of a forward particle from the fragments of the corresponding parent hadron. 
Using the BFKL jargon, these coefficients are also known as forward-production impact factors.
The Green’s function is governed by an integral evolution equation, whose kernel is known at the \ac{NLO} perturbative accuracy~\cite{Fadin:1998py,Ciafaloni:1998gs,Fadin:1998jv,Fadin:2000kx,Fadin:2000hu,Fadin:2004zq,Fadin:2005zj} (see Refs.~\cite{Caola:2021izf,Falcioni:2021dgr,DelDuca:2021vjq,Byrne:2022wzk,Fadin:2023roz,Byrne:2023nqx} for ongoing efforts in calculating higher-order corrections).

The predictive capability of the high-energy resummation from BFKL at NLL is constrained by the availability of off-shell emission functions computed within NLO. 
They include: 
a) colliding-parton (quarks and gluons) impact factors~\cite{Fadin:1999de,Fadin:1999df}, which are needed to calculate 
b) forward-jet~\cite{Bartels:2001ge,Bartels:2002yj,Caporale:2011cc,Caporale:2012ih,Ivanov:2012ms,Colferai:2015zfa} and 
c) forward light hadron~\cite{Ivanov:2012iv} emission functions. 
Additionally, we mention: 
d) virtual photon to light vector meson~\cite{Ivanov:2004pp}, 
e) light-by-light impact factors~\cite{Bartels:2000gt,Bartels:2001mv,Bartels:2002uz,Bartels:2004bi,Fadin:2001ap,Balitsky:2012bs}, and
f) the emission function for the production of a forward Higgs boson in the infinite top-mass limit~\cite{Hentschinski:2020tbi,Celiberto:2022fgx} (see also Refs.~\cite{Hentschinski:2022sko,Fucilla:2022whr}).
Staying at \ac{LO}, we include:
Drell--Yan pairs~\cite{Hentschinski:2012poz,Motyka:2014lya},
heavy-quark pairs~\cite{Celiberto:2017nyx,Bolognino:2019ccd,Bolognino:2019yls}, and 
forward $J/\psi$ emitted at low transverse momentum~\cite{Boussarie:2017oae} (see also Refs.~\cite{Boussarie:2015jar,Boussarie:2016gaq,Boussarie:2017xdy}).

Gold-plated phenomenological channels to probe high-energy QCD dynamics at hadron colliders essentially fall into the following classes: Mueller--Navelet dijets~\cite{Mueller:1986ey,Colferai:2010wu,Caporale:2012ih,Ducloue:2013hia,Ducloue:2013bva,Caporale:2013uva,Caporale:2014gpa,Colferai:2015zfa,Caporale:2015uva,Ducloue:2015jba,Celiberto:2015yba,Celiberto:2015mpa,Celiberto:2016ygs,Celiberto:2016vva,Caporale:2018qnm,deLeon:2021ecb,Celiberto:2022gji}, dihadron production~\cite{Celiberto:2016hae,Celiberto:2016zgb,Celiberto:2017ptm,Celiberto:2017uae,Celiberto:2017ydk} and multi-jet tags~\cite{Caporale:2015vya,Caporale:2015int,Caporale:2016soq,Caporale:2016vxt,Caporale:2016xku,Celiberto:2016vhn,Caporale:2016djm,Caporale:2016pqe,Chachamis:2016qct,Chachamis:2016lyi,Caporale:2016lnh,Caporale:2016zkc,Caporale:2017jqj,Chachamis:2017vfa}, hadron plus jet~\cite{Bolognino:2018oth,Bolognino:2019cac,Bolognino:2019yqj,Celiberto:2020wpk,Celiberto:2020rxb,Celiberto:2022kxx}, Higgs plus jet~\cite{Celiberto:2020tmb,Celiberto:2021fjf,Celiberto:2021tky,Celiberto:2021txb,Celiberto:2021xpm}, heavy-light dijet systems~\cite{Bolognino:2021mrc,Bolognino:2021hxx} and heavy-hadron~\cite{Boussarie:2017oae,Celiberto:2017nyx,Bolognino:2019ouc,Bolognino:2019yls,Bolognino:2019ccd,Celiberto:2021dzy,Celiberto:2021fdp,Bolognino:2022wgl,Celiberto:2022dyf,Celiberto:2022grc,Bolognino:2022paj,Celiberto:2022keu,Celiberto:2022zdg,Celiberto:2022kza,Celiberto:2024omj} emissions.

Remarkably, detections of single forward particles are excellent probe channels for the proton's content at low-$x$ through the BFKL \emph{Unintegrated Gluon Distribution} (UGD), whose energy evolution is ruled by the BFKL Green's function. 
We mention: light vector-meson leptoproduction at HERA~\cite{Anikin:2009bf,Anikin:2011sa,Besse:2013muy,Bolognino:2018rhb,Bolognino:2018mlw,Bolognino:2019bko,Bolognino:2019pba,Celiberto:2019slj,Luszczak:2022fkf} and the Electron-Ion Collider (EIC)~\cite{Bolognino:2021niq,Bolognino:2021gjm,Bolognino:2022uty,Celiberto:2022fam,Bolognino:2022ndh}, exclusive quarkonium photoproduction~\cite{Bautista:2016xnp,Garcia:2019tne,Hentschinski:2020yfm,Peredo:2023oym}, the inclusive detection of Drell--Yan dilepton systems~\cite{Motyka:2014lya,Brzeminski:2016lwh,Motyka:2016lta,Celiberto:2018muu} or bottomed jets~\cite{Chachamis:2015ona,Chachamis:2013bwa,Chachamis:2009ks}.

The information on the gluon content at small-$x$ provided by the BFKL UGD turned out to be decisive to improve the description of small-$x$ resummed collinear PDFs~\cite{Ball:2017otu,Abdolmaleki:2018jln,Bonvini:2019wxf}. 
Additionally, it connects with model studies of low-$x$ improved, twist-two gluon \ac{TMD} distributions~\cite{Bacchetta:2020vty,Celiberto:2021zww,Bacchetta:2021oht,Bacchetta:2021lvw,Bacchetta:2021twk,Bacchetta:2022esb,Bacchetta:2022crh,Bacchetta:2022nyv,Celiberto:2022omz,Bacchetta:2023zir,Bacchetta:2024fci}. 
Analyses presented in Refs.~\cite{Hentschinski:2021lsh,Mukherjee:2023snp} provide insights into the relationship between TMD and low-$x$ dynamics.
Studies in Refs.~\cite{Boroun:2023goy,Boroun:2023ldq} delve into the exploration of the connection between the UGD and color-dipole cross sections.

A striking challenge connected to the BFKL description of Mueller--Navelet rapidity distributions and azimuthal-angle correlations arises from the impact of NLL contributions. These NLL terms, while of the same order as the pure LL case, exhibit an opposite sign. This leads to strong instabilities within the resummed series, particularly when studies on \ac{MHOUs} via variations of energy scales around their natural values are made.

Consequently, Mueller--Navelet observables tend to assume unphysical values, especially as the rapidity separation between jets becomes sufficiently large. Likewise, azimuthal correlations display anomalous behaviors at both small and large rapidity distances. Various approaches have been explored to address this issue. Notably, the \ac{BLM} prescription~\cite{Brodsky:1996sg,Brodsky:1997sd,Brodsky:1998kn,Brodsky:2002ka}, specifically designed for semi-hard processes~\cite{Caporale:2015uva}, provides a partial mitigation of these instabilities in azimuthal correlations, leading to a moderate improvement in agreement with experimental data.

However, the efficacy of BLM is limited, particularly in the case of light dihadron or hadron plus jet semi-hard distributions. The primary reason for this limitation is that the optimal renormalization scale values recommended by BLM are significantly higher than the natural scales of the underlying processes~\cite{Celiberto:2017ius,Bolognino:2018oth,Celiberto:2020wpk}. Consequently, in these scenarios total cross sections suffer from a substantial reduction in statistics.

Compelling indications of a reached stabilization of the high-energy resummation under higher-order corrections and scale variations have been recently observed in the context of semi-hard reactions featuring final states sensitive to Higgs boson detections~\cite{Celiberto:2020tmb,Mohammed:2022gbk,Celiberto:2023rtu,Celiberto:2023uuk,Celiberto:2023eba,Celiberto:2023nym,Celiberto:2023dkr,Celiberto:2023rqp}.
A clear signature of this stabilizing trend came out for the first time from studies on the semi-inclusive emissions of $\Lambda_c$ hyperons~\cite{Celiberto:2021dzy} or singly bottomed hadrons~\cite{Celiberto:2021fdp} at the LHC. 
In particular, it was highlighted that the stabilizing effect is directly connected with the distinctive pattern exhibited by VFNS collinear FFs governing the production of these singly heavy-flavored particles at high transverse momentum.

Subsequent analyses on vector quarkonia~\cite{Celiberto:2022dyf,Celiberto:2023fzz}, charmed $B$ mesons~\cite{Celiberto:2022keu,Celiberto:2024omj}, and heavy-light tetraquarks~\cite{Celiberto:2023rzw}, clarified that this remarkable property, known as \emph{natural stability} of the high-energy resummation in QCD~\cite{Celiberto:2022grc}, emerges as an intrinsic feature inherently associated with final states sensitive to heavy flavor.

\subsection{Hybrid-factorization studies at $\NLL$ and beyond}
\label{ssec:hybrid_factorization}

\begin{figure*}[!t]
\centering

\includegraphics[width=0.475\textwidth]{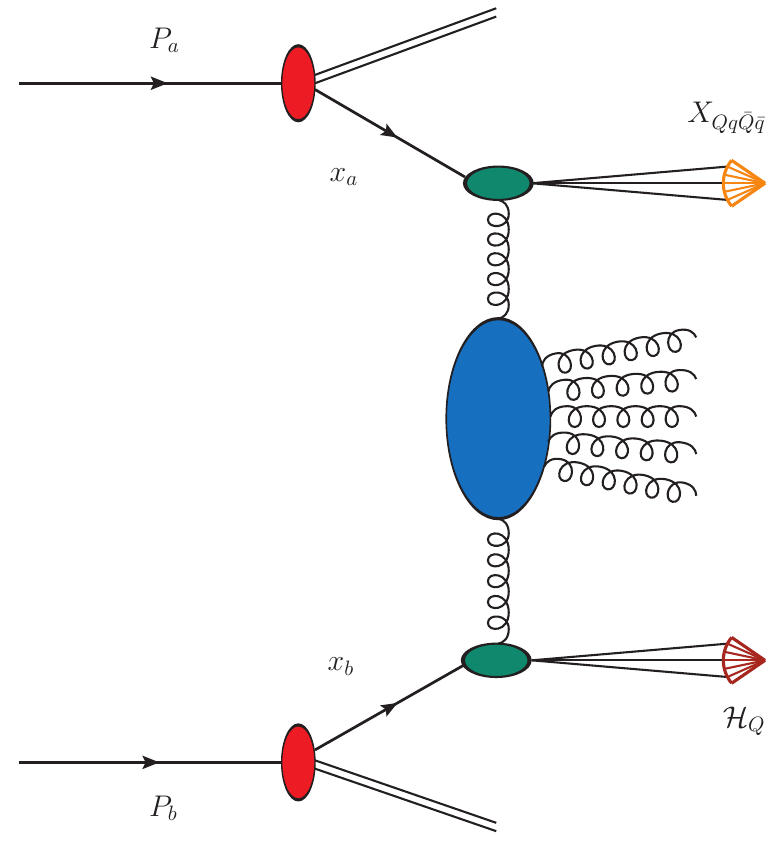}
   \hspace{0.50cm}
\includegraphics[width=0.475\textwidth]{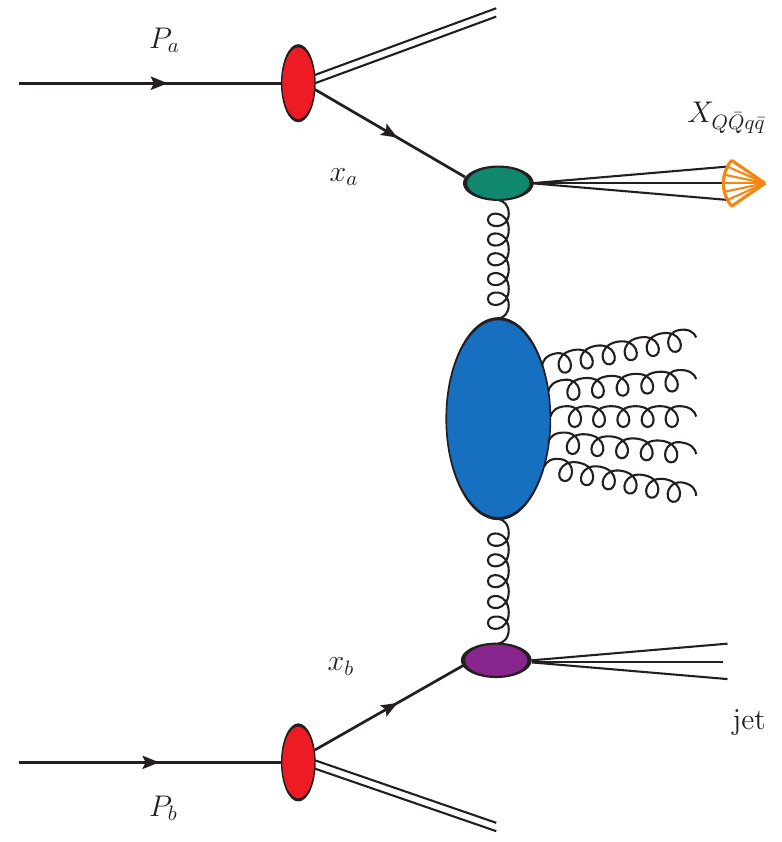}

\caption{Pictorial representation of the tetraquark~$+$~hadron (left) and tetraquark~$+$~jet (right) semi-inclusive hadroproduction within the hybrid collinear and high-energy factorization (figures realized with {\tt JaxoDraw 2.0}~\cite{Binosi:2008ig}). Red blobs depict collinear FFs. The off-shell hard factor, part of the hadron (jet) emission function, is given by green (violet) ovals. Tetraquark ($Q$-hadron) emissions are portrayed by orange (firebrick) arrows. The large blue blob at the center of each diagram represents the BFKL Green's function.}
\label{fig:process}
\end{figure*}

We investigate the two classes of processes represented in Fig.~\ref{fig:process}
\begin{equation}
\label{process}
\setlength{\jot}{10pt} 
\begin{split}
    {\rm p}(P_a) \;+\; {\rm p}(P_b) &\;\rightarrow\; \XQq(\bm{\kappa}_1, y_1) \;+\; {\cal X} \;+\; \HQ(\bm{\kappa}_2, y_2) \; ,
    \\
    {\rm p}(P_a) \;+\; {\rm p}(P_b) &\;\rightarrow\; \XQq(\bm{\kappa}_1, y_1) \;+\; {\cal X} \;+\; {\rm jet}(\bm{\kappa}_2, y_2) \; ,
\end{split}
\end{equation}
where a heavy-light tetraquark ($\Xcu$, $\Xcs$, $\Xbu$, or $\Xbs$) is emitted in association with singly heavy-flavored hadron ($\Hc$ or $\Hb$), or a light jet, ${\cal O} = \{ {\cal H}_c, {\cal H}_b, {\rm jet} \}$. 
The final-state particles possess large transverse momenta, $|\bm{\kappa}_{1,2}| \gg \Lambda_{\rm QCD}$, and their rapidity separation is $ \DY \equiv y_1 - y_2$. 

As for a ${\cal H}_c$ hadron, we refer to inclusive states consisting in the sum of fragmentation channels to single-charmed $D^\pm$, $D^0$ and $D^{*\pm}$~mesons, and also $\Lambda_c^\pm$ hyperons.
Analogously, a ${\cal H}_b$ particle stands as combinations of noncharmed $B$~mesons and $\Lambda_b^0$ baryons~\cite{Celiberto:2021fdp}.
The undetected gluon radiation is inclusively indicated as ${\cal X}$. 
Large observed transverse momenta and large rapidity distances are needed for dealing with semi-hard final-state configurations.
Furthermore, transverse-momentum ranges have to be large enough to make the validity of the VFNS collinear fragmentation be the dominant mechanism for the production of heavy hadrons.

Incoming protons' four-momenta can decomposed as Sudakov vectors satisfying $P_a^2= P_b^2=0$ and $2 (P_a\cdot P_b) = s$.
In this way, $\kappa_1$ and $\kappa_2$ can be cast as
\begin{equation}\label{sudakov}
\kappa_{1,2} = x_{1,2} P_{a,b} - \frac{\kappa_{1,2\perp}^{\,2}}{x_{1,2} s}P_{b,a} + \kappa_{1,2\perp} \ , \qquad
\bm{\kappa}_{1,2}^{\,2} \equiv -\kappa_{1,2\perp}^2\;.
\end{equation}
Final-state object's longitudinal fractions, $x_{1,2}$, depend on rapidities as
$y_{1,2}=\pm\frac{1}{2}\ln\frac{x_{1,2}^2 s}
{\bm{\kappa}_{1,2}^2}$.
Thus one have $\drv y_{1,2} = \pm \frac{\drv x_{1,2}}{x_{1,2}}$, and $\DY \equiv y_1 - y_2 = \ln\frac{x_1 x_2 s}{|\bm{\kappa}_1||\bm{\kappa}_2|}$.

The LO cross section of our reactions (Eq.~\eqref{process}) in pure collinear QCD would take the form of a one-dimensional convolution among protons' PDFs, hadrons' FFs, and partonic hard factors.
In the double hadron channel (panel a) of Fig.~\ref{fig:process}), one writes
\begin{equation}
\label{sigma_collinear_XH} 
\begin{split}
 \frac{\drv \sigma_{[p \;+\; p \;\;\to\;\; \XQq \;+\; \HQ]}^{\rm LO}}{\drv x_1 \drv x_2 \drv^2 \bm{\kappa}_1 \drv^2 \bm{\kappa}_2}
 &=\sum_{a,b} \int_0^1 \drv x_a \int_0^1 \drv x_b\ 
 f_a(x_a, \mu_F) f_b(x_b, \mu_F)
\\ 
 &\hspace{0.00cm}\times \, \int_{x_1}^1 \frac{\drv z_1}{z_1} \int_{x_2}^1 \frac{\drv z_2}{z_2}\
 D^{X}_a\left(\frac{x_1}{z_1}, \mu_F\right) D^{\cal H}_b\left(\frac{x_2}{z_2}, \mu_F\right)
 \frac{\drv {\hat\sigma}_{a,b}}
 {\drv x_a \drv x_b \drv z_1 \drv z_2 \drv^2 \bm{\kappa}_1 \drv^2 \bm{\kappa}_2}\;,
\end{split}
\end{equation}
where ($a,b$) indices run over quarks, antiquarks, and the gluon, $f_{a,b}$ are proton PDFs, $D^{X}_{a,b}$ $\big( D^{\cal H}_{a,b} \big)$ denote heavy-light tetraquark (singly heavy-flavored hadron) FFs, $x_{a,b}$ stand for the longitudinal fractions of the struck partons, $z_{1,2}$ are the longitudinal fractions of outgoing partons, and $\drv \hat\sigma_{a,b}$ are partonic-subprocess cross sections.

Analogously, in the tetraquark plus jet channel (panel b) of Fig.~\ref{fig:process}) we have
\begin{equation}
\label{sigma_collinear_XJ}
\begin{split}
 \frac{\drv \sigma_{[p \;+\; p \;\;\to\;\; {\XQq} \;+\; {\rm jet}]}^{\rm LO}}{\drv x_1 \drv x_2 \drv^2 \bm{\kappa}_1 \drv^2 \bm{\kappa}_2}
 &=\sum_{a,b} \int_0^1 \drv x_a \int_0^1 \drv x_b\ 
 f_a(x_a, \mu_F) f_b(x_b, \mu_F)
\int_{x_1}^1 \frac{\drv z}{z}D^{X}_{a}\left(\frac{x_1}{z}\right) 
\frac{\drv {\hat\sigma}_{a,b}}
{\drv x_1\drv x_2\drv z\,\drv ^2\bm{\kappa}_1\drv ^2\bm{\kappa}_2}\;.
\end{split}
\end{equation}

On the contrary, deriving the formula for the high-energy resummed cross section within our hybrid factorization requires a two-step process. 
First, we employ the high-energy factorization as prescribed by BFKL.
Then, we enhance the description by incorporating collinear components, PDFs and FFs. 
To this extent, we express the differential cross section as a Fourier sum of azimuthal-angle coefficients
\begin{equation}
 \label{dsigma_Fourier}
 \frac{\drv \sigma^\NLLp}{\drv y_1 \drv y_2 \drv \bm{\kappa}_1 \drv \bm{\kappa}_2 \drv \varphi_1 \drv \varphi_2} =
 \frac{1}{(2\pi)^2} \left[{\cal C}_0^\NLLp + 2 \sum_{n=1}^\infty \cos \left(n (\Phi - \pi)\right) \,
 {\cal C}_n^\NLLp \right]\, ,
\end{equation}
with $\varphi_{1,2}$ the observed azimuthal angles and $\Phi \equiv \varphi_1 - \varphi_2$.
Azimuthal coefficients are calculated within the BFKL formalism and they encode the LL and NLL resummation of high-energy logarithms. We rely upon the $\MSb$ renormalization scheme~\cite{PhysRevD.18.3998} to write (see Ref.~\cite{Caporale:2012ih} for details)
\begin{equation}
\label{Cn_NLLp_MSb}
\begin{split}
 \CnNLLp &= \int_0^{2\pi} \drv \varphi_1 \int_0^{2\pi} \drv \varphi_2\,
 \cos \left(n (\Phi - \pi)\right) \,
 \frac{\drv \sigma^\NLLp}{\drv y_1 \drv y_2\, \drv |\bm{\kappa}_1| \, \drv |\bm{\kappa}_2| \drv \varphi_1 \drv \varphi_2}\;
\\
 &= \; \frac{e^{\DY}}{s} 
 \int_{-\infty}^{+\infty} \drv \nu \, e^{{\DY} \bar \alpha_s(\mu_R)\chi^\NLO(n,\nu)}
\\
 &\times \; \alpha_s^2(\mu_R) \, 
 \biggl\{
 \F_1^\NLO(n,\nu,|\bm{\kappa}_1|, x_1)[\F_2^\NLO(n,\nu,|\bm{\kappa}_2|,x_2)]^*\,
\\ 
 &+ \,
 \left.
 \bar \alpha_s^2(\mu_R)
 \, \DY
 \frac{\beta_0}{4 N_c}\chi(n,\nu)f(\nu)
 \right\} \;,
\end{split}
\end{equation}
where $\bar \alpha_s(\mu_R) \equiv \alpha_s(\mu_R) N_c/\pi$ with $N_c$ the color number, $\beta_0 = 11N_c/3 - 2 n_f/3$ is the first coefficient of the QCD $\beta$-function with $n_f$ the flavor number.
We select a two-loop running coupling with initial condition $\alpha_s\left(M_Z\right)=0.11707$ and a dynamic $n_f$.
The BFKL kernel at the exponent of Eq.~\eqref{Cn_NLLp_MSb} reads
\begin{eqnarray}
 \label{chi}
 \chi^\NLO(n,\nu) = \chi(n,\nu) + \bar\alpha_s \hat \chi(n,\nu) \;,
\end{eqnarray}
with
\begin{eqnarray}
 \label{kernel_LO}
 \chi\left(n,\nu\right) = -2\gamma_{\rm E} - 2 \, {\rm Re} \left\{ \psi\left(\frac{1+n}{2} + i \nu \right) \right\} \, 
\end{eqnarray}
being the LO BFKL eigenvalues, $\gamma_{\rm E}$ the Euler-Mascheroni constant, and $\psi(z) \equiv \Gamma^\prime
(z)/\Gamma(z)$ the logarithmic derivative of the Gamma function. The $\hat\chi(n,\nu)$ function in Eq.~\eqref{chi} is the NLO kernel correction
\begin{equation}
\begin{split}
\label{chi_NLO}
\hat \chi\left(n,\nu\right) &= \bar\chi(n,\nu)+\frac{\beta_0}{8 N_c}\chi(n,\nu)
\left(-\chi(n,\nu)+10/3+2\ln\frac{\mu_R^2}{\mu_C^2}\right) \;,
\end{split}
\end{equation}

with $\mu_C \equiv \sqrt{m_{1 \perp} m_{2 \perp}}$, where $m_{(1,2) \perp}$ are the observed-particle transverse masses.
Masses or our heavy-light tetraquark are set to $m_X = 2(m_q+m_Q)$, with $M_q$ ($m_Q$) being the mass of the light (heavy) constituent quark.
The transverse mass of the light-flavored jet merely coincides with its transverse momentum, $|\bm{\kappa}_J|$.
The characteristic $\bar\chi(n,\nu)$ function was obtained in Ref.~\cite{Kotikov:2000pm}
\begin{equation}
 \label{kernel_NLO}
 \bar \chi(n,\nu)\,=\, - \frac{1}{4}\left\{\frac{\pi^2 - 4}{3}\chi(n,\nu) - 6\zeta(3) - \frac{\drv^2 \chi}{\drv\nu^2} + \,2\,\phi(n,\nu) + \,2\,\phi(n,-\nu)
 \right.
\end{equation}
\[
 \left.
 +\; \frac{\pi^2\sinh(\pi\nu)}{2\,\nu\, \cosh^2(\pi\nu)}
 \left[
 \left(3+\left(1+\frac{n_f}{N_c^3}\right)\frac{11+12\nu^2}{16(1+\nu^2)}\right)
 \delta_{n0}
 -\left(1+\frac{n_f}{N_c^3}\right)\frac{1+4\nu^2}{32(1+\nu^2)}\delta_{n2}
\right]\right\} \, ,
\]
with
\begin{equation}
\label{kernel_NLO_phi}
 \phi(n,\nu)\,=\,-\int_0^1 \drv x\,\frac{x^{-1/2+i\nu+n/2}}{1+x}\left\{\frac{1}{2}\left(\psi^\prime\left(\frac{n+1}{2}\right)-\zeta(2)\right)+\mbox{Li}_2(x)+\mbox{Li}_2(-x)\right.
\end{equation}
\[
\left.
 +\; \ln x\left[\psi(n+1)-\psi(1)+\ln(1+x)+\sum_{k=1}^\infty\frac{(-x)^k}{k+n}\right]+\sum_{k=1}^\infty\frac{x^k}{(k+n)^2}\left[1-(-1)^k\right]\right\}
\]
\[
 =\; \sum_{k=0}^\infty\frac{(-1)^{k+1}}{k+(n+1)/2+i\nu}\left\{\psi^\prime(k+n+1)-\psi^\prime(k+1)\right.
\]
\[
 \left.
 +\; (-1)^{k+1}\left[\beta_{\psi}(k+n+1)+\beta_{\psi}(k+1)\right]-\frac{\psi(k+n+1)-\psi(k+1)}{k+(n+1)/2+i\nu}\right\} \; ,
\]
where
\begin{equation}
\label{kernel_NLO_phi_beta_psi}
 \beta_{\psi}(z)=\frac{1}{4}\left[\psi^\prime\left(\frac{z+1}{2}\right)
 -\psi^\prime\left(\frac{z}{2}\right)\right] \; ,
\end{equation}
and
\begin{equation}
\label{dilog}
\mbox{Li}_2(x) = \int^x_0 \drv \omega \,\frac{\ln(1-\omega)}{\omega} \; .
\end{equation}
The singly off-shell emissions functions
\begin{equation}
\label{IFs}
\F_{1,2}^\NLO(n,\nu,|\bm{\kappa}_{1,2}|,x_{1,2}) =
\F_{1,2}(n,\nu,|\bm{\kappa}_{1,2}|,x_{1,2}) +
\alpha_s(\mu_R) \, \hat \F_{1,2}(n,\nu,|\bm{\kappa}_{1,2}|,x_{1,2})
\end{equation}
The LO expressions for these functions depicting the production of a forward hadron and a forward jet respectively read as
\begin{equation}
\label{LOHEF}
\begin{split}
\F_h(n,\nu,|\bm{\kappa}_h|,x_h) 
&= 2 \, \sqrt{\frac{C_F}{C_A}}
|\bm{\kappa}_h|^{2i\nu-1}\,\int_{x_h}^1\frac{\drv z}{z}
\left(\frac{z}{x_h} \right)
^{2 i\nu-1} 
 \left[\frac{C_A}{C_F}f_g(z)D_g^h\left(\frac{x_h}{z}\right)
 +\sum_{a=q,\bar q}f_a(z)D_a^h\left(\frac{x_h}{z}\right)\right] 
\end{split}
\end{equation}
and
\begin{equation}
 \label{LOJEF}
 \F_J(n,\nu,|\bm{\kappa}_J|,x_J) =  2 \sqrt{\frac{C_F}{C_A}}
 |\bm{\kappa}_J|^{2i\nu-1}\,\left[\frac{C_A}{C_F}f_g(x_J)
 +\sum_{b=q,\bar q}f_b(x_J)\right] \;,
\end{equation}
with $C_F \equiv (N_c^2-1)/(2N_c)$ and $C_A \equiv N_c$ the usual Casimir QCD factors.
The $f(\nu)$ function is related to a logarithmic derivative of LO emission functions
\begin{equation}
 f(\nu) = \frac{i}{2} \, \frac{\drv}{\drv \nu} \ln\left(\frac{\F_1}{\F_2^*}\right) + \ln\left(|\bm{\kappa}_1| |\bm{\kappa}_2|\right) \;.
\label{fnu}
\end{equation}

In Eq.~(\ref{Cn_NLLp_MSb}), the remaining components are the NLO emission-function corrections, denoted as $\hat \F_{1,2}$. 
The forward-hadron NLO term is determined under the was calculated in Ref.~\cite{Ivanov:2012iv} and its expression is provided in Appendix~\hyperlink{app:NLOHEF}{A} of this review. 
As for the forward-jet NLO term, our choice follows Refs.~\cite{Ivanov:2012iv,Ivanov:2012ms}. 
To ease numerical analyses, we employ a jet selection function\footnote{We remind the reader that the most popular jet-reconstruction functions fall into two major classes (see Refs.~\cite{Chekanov:2002rq,Salam:2010nqg} and Refs. therein): \emph{cone-type} and \emph{sequential-clustering} algorithms (such as the well known (\emph{anti-})$\kappa_\perp$ selection function~\cite{Catani:1993hr,Cacciari:2008gp}).} calculated within the \ac{SCA}~\cite{Furman:1981kf,Aversa:1988vb} in its cone-type version~\cite{Colferai:2015zfa} and with the jet cone radius set to ${\cal R} = 0.5$. The analytic expression for this emission function is given in Appendix~\hyperlink{app:NLOJEF}{B}.

A proper way for a phenomenological comparison between our hybrid factorization and pure fixed-order results would necessitate for a numerical framework tailored for computing NLO distributions two-particle reactions in hadron collisions.
According to our knowledge, such a technology is currently unavailable.
For the sake of comparison with reference fixed-order predictions, we truncate the expansion of azimuthal coefficients of Eq.~\eqref{Cn_NLLp_MSb} up to the ${\cal O}(\alpha_s^3)$ levels. 
This gives us an effective high-energy fixed-order ($\HENLOp$) expression, 
which captures the leading-power asymptotic signal present a pure NLO calculation and, at the same time, disregards terms proportional to inverse powers of the partonic center-of-mass energy.
The $\MSb$ expressions for the azimuthal coefficients at $\HENLOp$ reads
\begin{equation}
\label{Cn_HENLO_MSb}
 \CnHENLOp =
 \frac{e^{\DY}}{s}
 \int_{-\infty}^{+\infty} \drv \nu \,
 \alpha_s^2(\mu_R) \,
 \left[ 1 + \bar \alpha_s(\mu_R) \DY \chi(n,\nu) \right] \,
 \F_1^\NLO(n,\nu,|\bm{\kappa}_1|, x_1) \,[\F_2^\NLO(n,\nu,|\bm{\kappa}_2|,x_2)]^*
 \;,
\end{equation}
with the exponentiated kernel expanded and truncated at ${\cal O}(\alpha_s)$.
Moreover, we present predictions at a pure LL order, given by
\begin{equation}
\label{Cn_LL_MSb}
  \CnLL = \frac{e^{\DY}}{s} 
 \int_{-\infty}^{+\infty} \drv \nu \, e^{{\DY} \bar \alpha_s(\mu_R)\chi(n,\nu)} \, \alpha_s^2(\mu_R) \, \F_1(n,\nu,|\bm{\kappa}_1|, x_1)[\F_2(n,\nu,|\bm{\kappa}_2|,x_2)]^* \,.
\end{equation}

Eqs.~(\ref{Cn_NLLp_MSb}) to~(\ref{Cn_LL_MSb}) tells us the way our hybrid factorization is constructed. According to BFKL, the hadronic cross section is high-energy factorized as a transverse-momentum convolution between the Green's function and the two off-shell emission functions. 
These impact factors embody collinear PDFs and FFs.
The $\NLLp$ label emphasizes the complete resummation of energy logarithms at NLL accuracy by means of perturbative ingredients calculated at NLO. 
The `$+$' superscript in Eq.~(\ref{Cn_NLLp_MSb}) highlights the fact that some next-to-NLL contributions, resulting from the cross product of the two NLO impact-factor corrections, are also accounted for.

Renormalization ($\mu_R$) and factorization ($\mu_F$) scales are set to the \emph{natural} energies suggested by process kinematics. 
Thus, we have $\mu_R = \mu_F = \mu_N = m_{1 \perp} + m_{2 \perp}$.
As for collinear PDFs, we make use of the {\tt NNPDF4.0} NLO set~\cite{NNPDF:2021uiq,NNPDF:2021njg} as implemented in the {\tt LHAPDF~v6.5.4} interface~\cite{Buckley:2014ana}.
Such PDFs were extracted by means of global fits and through the so-called \emph{replica} method, originally derived in Ref.~\cite{Forte:2002fg} in the context of neural-network techniques and now widely employed on multi-dimensional analyses of the proton structure~\cite{Bacchetta:2017gcc,Scimemi:2019cmh,Bacchetta:2019sam,Bacchetta:2022awv,Bury:2022czx,Moos:2023yfa} (see Ref.~\cite{Ball:2021dab} for a quantitative study on ambiguities emerging from \emph{correlations} among different PDF sets).
All calculations presented in this review are performed in the $\MSb$ renormalization scheme~\cite{PhysRevD.18.3998}.

\section{Heavy-flavor fragmentation: From heavy-light hadrons to tetraquarks}
\label{sec:HF_fragmentation}

In this Section we present our strategy to depict inclusive emissions heavy-flavored hadrons via the VFNS collinear fragmentation.
Section~\ref{ssec:natural_stability} is for a digression on the emergence of the \emph{natural stability}~\cite{Celiberto:2022grc}.
Features of the novel {\tt TQHL1.0} FF determinations for heavy-light tetraquark state are discussed in detail in Section~\ref{ssec:TQHL10_FFs}.

\subsection{Rise and discovery of the natural stability}
\label{ssec:natural_stability}

The direct link between the dynamics governing DGLAP-evolving VFNS FFs and the stabilization pattern of the hybrid factorization was initially uncovered via studies of semi-hard emissions of singly heavy-flavored hadrons, including $D$ mesons~\cite{Kniehl:2004fy,Kniehl:2005de,Kniehl:2006mw,Kneesch:2007ey,Corcella:2007tg,Anderle:2017cgl,Salajegheh:2019nea,Salajegheh:2019srg,Soleymaninia:2017xhc}, $\Lambda_c$ hyperons~\cite{Kniehl:2005de,Kniehl:2020szu,Delpasand:2020vlb}, and $b$-flavored (${\cal H}_b$) hadrons~\cite{Binnewies:1998vm,Kniehl:2007erq,Kniehl:2008zza,Kniehl:2011bk,Kniehl:2012mn,Kramer:2018vde,Kramer:2018rgb,Salajegheh:2019ach,Kniehl:2021qep}.

A surprising and unexpected stabilizing trend under MHOU studies came out in observables sensitive to the forward semi-inclusive emission of $\Lambda_c$\cite{Celiberto:2021dzy,Celiberto:2022rfj,Bolognino:2022wgl}, $D^{* \pm}$\cite{Celiberto:2022zdg,Bolognino:2022paj}, and ${\cal H}_b$\cite{Celiberto:2021fdp,Celiberto:2022rfj} particles. 
It was then confirmed by analyses of vector quarkonia and charmed $B$ mesons produced via the single-parton (leading-twist) fragmentation mechanism relying upon an initial-scale input~\cite{Braaten:1993mp,Zheng:2019dfk,Braaten:1993rw,Chang:1992bb,Braaten:1993jn,Ma:1994zt,Zheng:2019gnb,Zheng:2021sdo,Feng:2021qjm}. 
Contextually, novel determinations of DGLAP-evolving FFs for vector quarkonia and $B_c^{(*)}$ mesons were obtained in Refs.~\cite{Celiberto:2022dyf,Celiberto:2022kza} and~\cite{Celiberto:2022keu}, respectively.
A weaker, but still present stabilization trend also rises when $s$-flavored, cascade $\Xi^-/\bar\Xi^+$ baryons are detected~\cite{Celiberto:2022kxx}.

The main outcome of Refs.~\cite{Celiberto:2022grc,Celiberto:2021dzy,Celiberto:2021fdp,Celiberto:2022dyf,Celiberto:2022keu} was a clear indication that gluon collinear FF channel has a crucial role in our NLL hybrid factorization framework. 
Its energy dependence is responsible for the stability of the high-energy logarithmic series in our observables. 
Specifically, in the kinematic sectors of interest, where $10^{-4} \lesssim x \lesssim 10^{-2}$, the gluon PDF dominates over all (anti)quark channels. 
Since the gluon FF is convoluted diagonally with the gluon PDF in the LO hadron impact factor (see Eq.(\ref{LOHEF})), its behavior is significantly amplified. 
This characteristic persists even at NLO~\cite{Celiberto:2021fdp}, when the $(qg)$ and $(gq)$ nondiagonal channels are active (see Appendix\hyperlink{app:NLOHEF}{A}).

While the QCD running coupling decreases with $\mu_R$, and this affects both the Green's function and the impact factors, it is well-known that the gluon PDF grows with $\mu_F$.
When the latter is convoluted in the emission function with a gluon FF that also increases with $\mu_F$, as it happens for heavy-flavored hadrons, these two effects counteract each other. The net results is a stabilizing trend of heavy-hadron distributions under MHOU studies. 
The more pronounced the growth with $\mu_F$ of the gluon FF, the clearer the stabilization pattern becomes. 
This is the reason why distributions sensitive to singly bottom-flavored states are generally more stable charm-sensitive ones~\cite{Celiberto:2021fdp}. 
In contrast, when the gluon FF decreases with $\mu_F$, as observed for lighter hadron species~\cite{Celiberto:2021dzy,Celiberto:2022kxx,Celiberto:2024omj}, no stabilization under MHOU is evident. 
This hampers any possibility of conducting precision studies of high-energy distributions at natural energy scales~\cite{Celiberto:2020wpk}.

The manifestation of \emph{natural stability} in the presence of both singly heavy-flavored hadrons or quarkonia clearly highlights that this remarkable property is an \emph{intrinsic} characteristic of heavy-flavor emissions. 
It becomes apparent whenever a heavy-hadron species is detected and it does not depend on the basis assumptions made to build corresponding collinear FFs.

\subsection{The {\tt TQHL1.0} FF determinations}
\label{ssec:TQHL10_FFs}

Here we present our {\tt TQHL1.0} functions.
They are VFNS, DGLAP-evolved collinear FFs describing the direct inclusive production of a $S$-wave $\XQq$ tetraquark state within the single-parton, leading-twist fragmentation mechanism.
We essentially follow a two-step strategy. First, we define the initial energy-scale input for our FFs.
Then, we obtain phenomenology-ready FF determinations released as {\tt LHAPDF} grids.

\begin{figure*}[!t]
\centering
\includegraphics[width=0.45\textwidth]{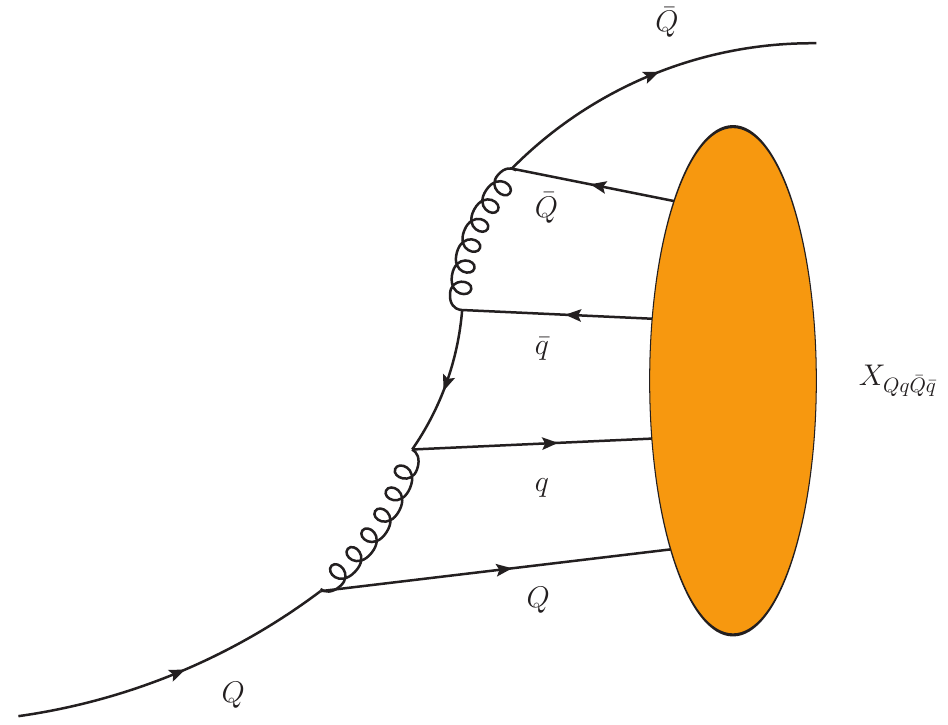}

\caption{Leading diagram for the fragmentation of a heavy quark $Q$ into a $\XQq$ tetraquark.
The orange blob portrays the nonperturbative hadronization of the $(Qq\bar{Q}\bar{q})$ system into the bound state.
The diagram was made by using {\tt JaxoDraw 2.0}~\cite{Binosi:2008ig}.}
\label{fig:FF_diagram}
\end{figure*}

Our approach to construct a tetraquark FF set begins with the calculation of the $(Q \to \XQq)$ $S$-wave collinear function, as carried out in Ref.~\cite{Nejad:2021mmp} (see Fig.~\ref{fig:FF_diagram}). This computation relies on the spin-dependent Suzuki model~\cite{Suzuki:1977km,Suzuki:1985up}, which accounts for transverse-momentum dependence. The collinear limit is obtained by neglecting the relative motion of constituent quarks inside the tetraquark~\cite{Lepage:1980fj,Brodsky:1985cr,Amiri:1986zv}.

The treatment of initial-scale input for tetraquark fragmentation follows a similar factorization scheme as that for quarkonia in \ac{NRQCD}~\cite{Caswell:1985ui,Thacker:1990bm,Bodwin:1994jh,Cho:1995vh,Cho:1995ce,Leibovich:1996pa,Bodwin:2005hm}.
There, a constituent $(Q\bar{Q})$ pair is produced perturbatively, after which tetraquark formation occurs via nonperturbative long-distance matrix elements.
In our formulation, a four-quark $(Qq\bar{Q}q)$ system is first emitted through perturbative splittings. 
Subsequently, its production amplitude is convoluted with a bound-state wave function that encapsulates the non-perturbative dynamics of tetraquark formation, according to the Suzuki model.

Starting from the heavy-quark input in Fig.~\ref{fig:FF_diagram}, taken at the initial scale of ${\cal Q}_0 = m_Q + m_X$, we generate our DGLAP-evolved set of collinear FFs for $S$-wave $\XQq$ tetraquarks.
The given ${\cal Q}_0$ value is nothing but the minimum required energy to produce the $(Qq\bar{Q}\bar{q})$ system in a color-singlet configuration.
Several tools, such as {\tt QCD-PEGASUS}~\cite{Vogt:2004ns}, {\tt HOPPET}~\cite{Salam:2008qg}, {\tt QCDNUM}~\cite{Botje:2010ay}, {\tt APFEL(++)}~\cite{Bertone:2013vaa,Carrazza:2014gfa,Bertone:2017gds}, and {\tt EKO}~\cite{Candido:2022tld}, come as public tools suited to numerically solve the DGLAP equations. 
Contrariwise to collinear PDFs, whose evolution is space-like, FF DGLAP evolution is time-like~\cite{Curci:1980uw,Furmanski:1980cm}. 
In this work we make use of {\tt APFEL++} and we set the evolution accuracy at NLO.

Light partons and nonconstituent heavy-quark channels are obtained through the DGLAP evolution. 
Thus, for each $\XQq$ species, we obtain a phenomenology-ready FF determination in {\tt LHAPDF} format, named \emph{TetraQuarks with Heavy and Light flavors} (\texttt{TQHL1.0}) functions.

It could be argued that our methodology overlooks the initial-scale contribution of light partons and nonconstituent heavy quarks, which are only generated through evolution at scales $\mu_F > {\cal Q}_0$. However, as highlighted in Ref.~\cite{Nejad:2021mmp}, these channels are deemed negligible at the initial scale ${\cal Q}_0$. This observation holds true for vector-quarkonium FFs as well, as discussed in Ref.~\cite{Celiberto:2022dyf}.

\begin{figure*}[!t]
\centering

   \includegraphics[scale=0.50,clip]{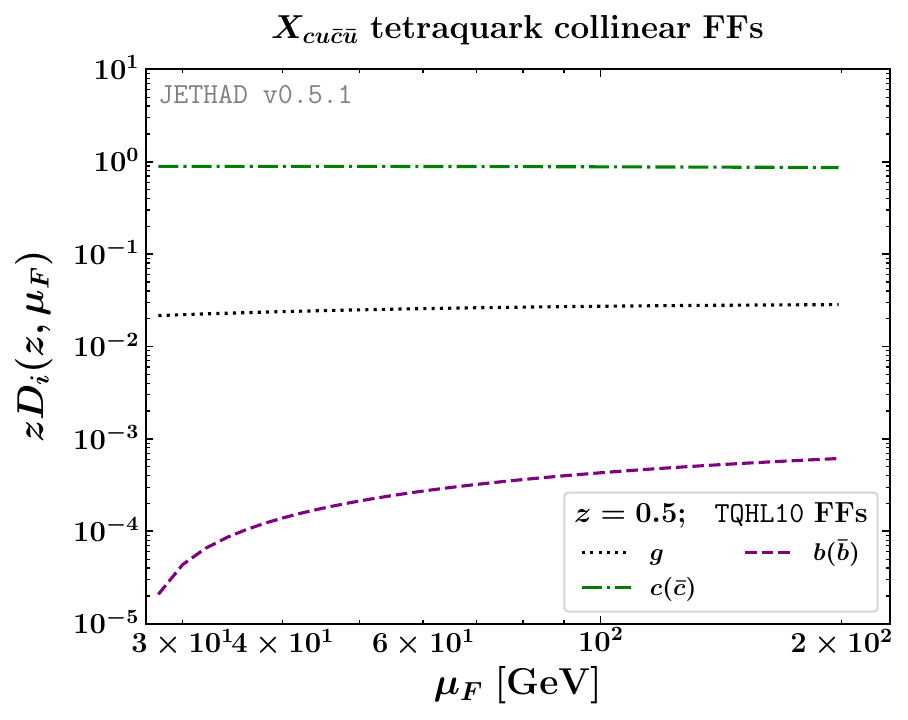}
   \includegraphics[scale=0.50,clip]{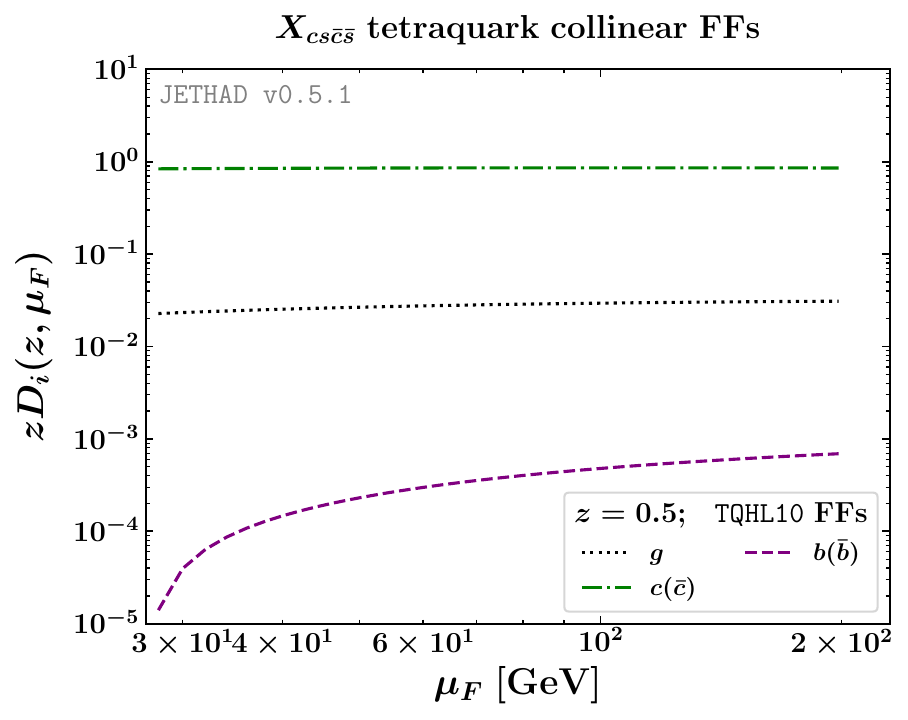}

   \includegraphics[scale=0.50,clip]{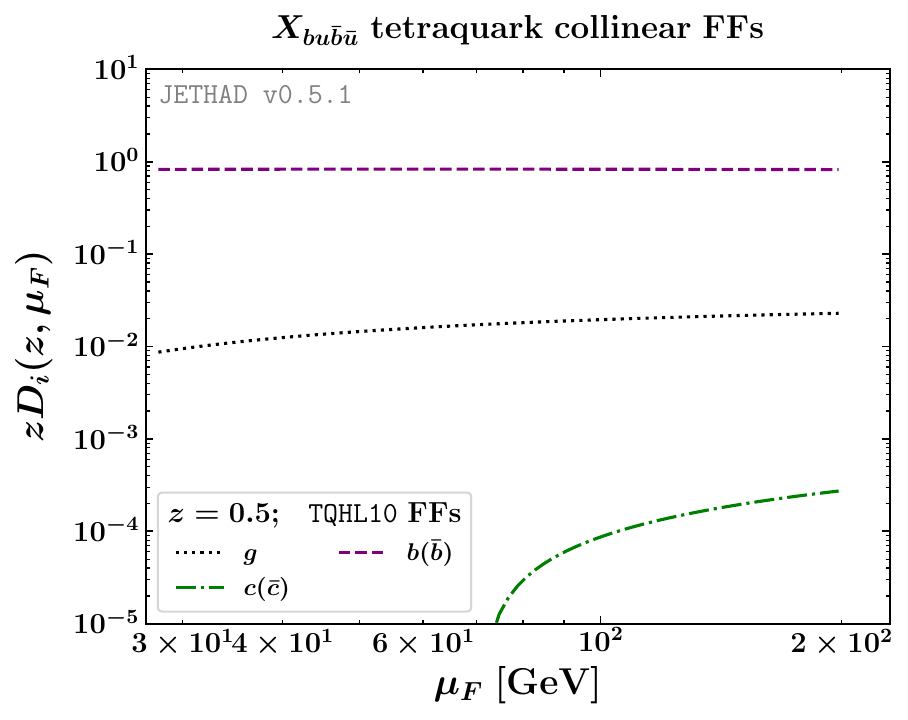}
   \includegraphics[scale=0.50,clip]{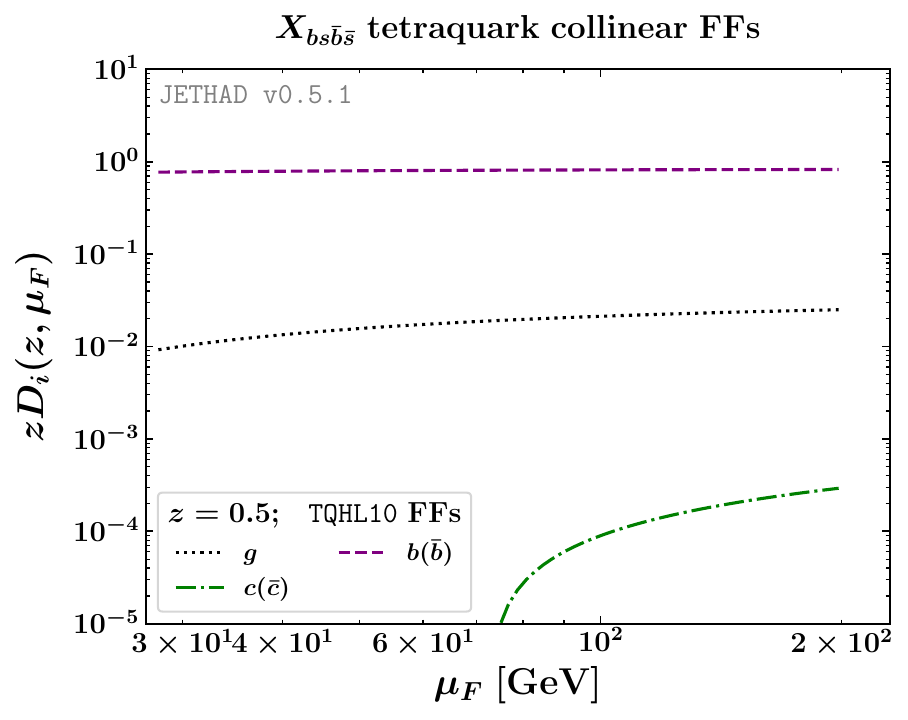}

\caption{Dependence on $\mu_F$ of the four {\tt TQHL1.0} collinear FF sets describing $\XQq$ tetraquark formation at $z \simeq \langle z \rangle = 0.5$.}
\label{fig:NLO_FFs_XQq}
\end{figure*}

Panels of Fig.~\ref{fig:NLO_FFs_XQq} is to show the dependence on $\mu_F$ of the four {\tt TQHL1.0} collinear FF sets describing $\XQq$ tetraquark formation at momentum fraction $z = 0.5$, which roughly represent its average value, $\langle z \rangle$.
As expected, the constituent heavy-quark FF channel heavily prevail over gluon and nonconstituent heavy-quark ones.
The other light-quark channels are not shown, since the are almost negligible.
Remarkably, the gluon FF smoothly increases with energy.
This supports the statement that \emph{natural stability} is encoded also in the $\XQq$ fragmentation mechanism.

\section{Exotic tetraquarks at the HL-LHC with {\Jethad}}
\label{sec:phenomenology}

All the predictions given in this review were obtained by making use of {\Jethad}, a hybrid and multimodular interface which combines both \textsc{Python}- and \textsc{Fortran}-based modules. 
{\Jethad} is designed for the computation, management, and processing of physical distributions defined within various formalisms~\cite{Celiberto:2020wpk,Celiberto:2022rfj,Celiberto:2023fzz}. 
In particular, numeric calculations of differential distributions were carried out via some of the \textsc{Fortran 2008} modular routines within {\Jethad}, whereas the native \textsc{Python 3.0} analyzer served as a reference tool for final elaborations.

Section~\ref{ssec:jethad} provides an overview of the key features of the current version of {\Jethad} technology ({\tt v0.5.1}), which is not yet public.
Further details on our error analysis are outlined in Section~\ref{ssec:uncertainty}. 
Information regarding final-state kinematic cuts enforced can be found in Section~\ref{ssec:final_state}.
Numeric results and discussions on rapidity-interval and transverse-momentum rates are presented in Sections~\ref{ssec:Y_rates} and~\ref{ssec:pT_rates}, respectively.

\subsection{The {\Jethad} {\tt v0.5.1} multimodular interface}
\label{ssec:jethad}

The beginning of the {\Jethad} project traces back to late 2017, driven by the need for accurate predictions of semi-hard hadron~\cite{Celiberto:2016hae,Celiberto:2017ptm} and jet~\cite{Celiberto:2015yba,Celiberto:2016ygs,Bolognino:2018oth} sensitive final states at the LHC.
Phenomenological studies of such reactions, proposed as probe channels for the high-energy resummation in QCD, necessitated the establishment of a reference numeric technology devoted to the computation and analysis of high-energy related distributions.

{\Jethad} {\tt v0.2.7} was the first named version and it played a key role in providing us with a pioneering BFKL-versus-DGLAP analysis in the context of semi-inclusive hadron-plus-jet emissions at the LHC~\cite{Celiberto:2020wpk}. 
Subsequent versions introduced new features, such as selecting forward heavy-quark pair observables ({\tt v0.3.0}~\cite{Bolognino:2019yls}), enabling studies on Higgs emissions and transverse-momentum distributions ({\tt v0.4.2}~\cite{Celiberto:2020tmb}), and integrating the \textsc{Python} analyzer with the \textsc{Fortran} core supermodule ({\tt v0.4.3}~\cite{Bolognino:2021mrc}).

Advancements went on with the possibility to make analyses on heavy-flavored hadrons via VFNS FFs at NLO ({\tt v0.4.4}~\cite{Celiberto:2021dzy}). 
The \deffont{\rlap{D}Unamis} (\textsc{DYnamis}) work package, dedicated to the forward Drell--Yan dilepton reaction~\cite{Celiberto:2018muu}, became part of {\Jethad} in {\tt v0.4.5}. 
The integration with the \emph{Leptonic-Exclusive-Amplitudes} (\textsc{LExA}) modular code allowed {\Jethad} to explore the proton content at low-$x$ through small-$x$ TMD densities in {\tt v0.4.6}~\cite{Bolognino:2021niq}.

Version {\tt v0.4.7}~\cite{Celiberto:2022dyf} introduced quarkonium-sensitive reactions from NRQCD leading-twist fragmentation. 
The latest features in {\tt v0.5.0}~\cite{Celiberto:2023fzz} and {\tt v0.5.1}~\cite{Celiberto:2024omj} encompass an enhanced system for MHOU-related studies, an expanded list of observables with a focus on singly- and doubly-differential transverse-momentum production rates~\cite{Celiberto:2022gji,Celiberto:2022kxx,Celiberto:2024omj}, and support for \emph{matching} procedures with collinear factorization~\cite{Celiberto:2023rtu,Celiberto:2023uuk,Celiberto:2023eba,Celiberto:2023nym,Celiberto:2023rqp}.

From the fundamental core to service modules and routines, {\Jethad} has been designed to dynamically achieve high levels of computational performance. 
The multidimensional integrators within {\Jethad} leverage extensive parallel computing to actively choose the most suitable integration algorithm based on the shape of the integrand.

Any reaction analyzable with {\Jethad} can be dynamically selected through an intuitive, \emph{structure} based smart-management interface.
Physical final-state particles are represented by \emph{object} prototypes within this interface, where particle objects encapsulate all pertinent information about their physical counterparts, ranging from mass and charge to kinematic ranges and rapidity tags. 
These particle objects are initially loaded from a master database using a dedicated \emph{particle generation} routine, and custom particle generation is also supported. 
Then, these objects are \emph{cloned} into a final-state vector and \emph{injected} from the integrand routine to the corresponding, process-specific module by a dedicated \emph{controller}.

The flexibility in generating the physical final states is accompanied by a range of options for selecting the initial state. 
A unique \emph{particle-ascendancy} structure attribute enables {\Jethad} to rapidly learn whether a object is hadroproduced, electroproduced, photoproduced, etc.
This dynamic feature ensures that only relevant modules are initialized, optimizing computing-time efficiency.

{\Jethad} is structured as an \emph{object-based} interface that is entirely independent of the specific reaction under investigation. 
While originally inspired by high-energy QCD and TMD factorization phenomenology, the code's design allows for easy encoding of different approaches by simply implementing novel, dedicated (super)modules. 
These can be straightforwardly linked to the core structure of the code by means of a natively-equipped \emph{point-to-routine} system, making {\Jethad} a versatile, particle-physics oriented environment.

Having in mind providing the Scientific Community with a standard computation technology tailored for the management of diverse processes (described by distinct formalisms), we envision releasing the first public version of {\Jethad} in the medium-term future.

\subsection{Error analysis}
\label{ssec:uncertainty}

A commonly employed methodology for gauging the impact of MHOUs involves evaluating the sensitivity of our observables to variations of the renormalization scale and the factorization one around their natural values.

It is widely recognized that MHOUs strongly contribute to the overall uncertainty~\cite{Celiberto:2022rfj}. To assess their weight, we vary $\mu_R$ and $\mu_F$ simultaneously around $\mu_N/2$ and $2 \mu_N$, with the $C_{\mu}$ parameter in the figures of Sections~\ref{ssec:Y_rates} and~\ref{ssec:pT_rates} given as $C_\mu \equiv \mu_{F}/\mu_N = \mu_{R}/\mu_N$.

Another potential source of uncertainty lies in proton PDFs. Recent analyses on high-energy production rates indicate that choosing different PDF parametrizations and members within the same set has a minimal effect~\cite{Bolognino:2018oth,Celiberto:2020wpk,Celiberto:2021fdp,Celiberto:2022rfj}. Thus, our observables will be calculated by considering just the central member of the {\tt NNPDF4.0} parametrization.

Additional uncertainties may arise from a \emph{collinear improvement} of the NLO kernel, involving the inclusion of renormalization-group (RG) terms to make the BFKL equation compatible with the DGLAP one in the collinear limit, or from changes in the renormalization scheme~\cite{Salam:1998tj,Ciafaloni:2003rd,Ciafaloni:2003ek,Ciafaloni:2000cb,Ciafaloni:1999yw,Ciafaloni:1998iv,SabioVera:2005tiv}. 
The impact of collinear-improvement techniques on semi-hard rapidity-differential rates is found to be contained within error bands produced by MHOUs~\cite{Celiberto:2022rfj}.

Then, the $\MSb$~\cite{PhysRevD.18.3998} to MOM~\cite{Barbieri:1979be,PhysRevLett.42.1435} renormalization-scheme transition was estimated in Ref.~\cite{Celiberto:2022rfj} and leads to systematically higher MOM results for rapidity distributions. However, these results remain within the MHOUs bands. We note, however, that a proper MOM analysis should be based on MOM-evolved PDFs and FFs, which are not currently available.

To derive uncertainty bands for our distributions, we combine MHOUs with the numerical errors generated by multidimensional integration (see Section~\ref{ssec:final_state}). The latter is consistently maintained below $1\%$ thanks to the {\Jethad} integrators.

\subsection{Final-state kinematic cuts}
\label{ssec:final_state}

The first observable matter of our investigation is the rapidity-interval rate, also known as $\DY$-distribution.
It is given by the ${\cal C}_0$ azimuthal coefficient, defined in Section~\ref{ssec:hybrid_factorization}, integrated over transverse momenta and rapidities of the two outgoing particles, while their rapidity distance, $\DY$, is kept fixed.
We write
\begin{equation}
 \label{DY_rate}
 \frac{\drv \sigma(\DY, s)}{\drv \DY} =
 \int_{|\bm{\kappa}_1|^{\rm min}}^{|\bm{\kappa}_1|^{\rm max}} 
 \!\!\drv |\bm{\kappa}_1|
 \int_{|\bm{\kappa}_2|^{\rm min}}^{|\bm{\kappa}_2|^{\rm max}} 
 \!\!\drv |\bm{\kappa}_2|
 \int_{\max \, (y_1^{\rm min}, \, y_2^{\rm min} + \DY)}^{\min \, (y_1^{\rm max}, \, y_2^{\rm max} + \DY)} \drv y_1
 \, \,
 {\cal C}_0^{\rm [accuracy]}\left(|\bm{\kappa}_1|, |\bm{\kappa}_2|, y_1, y_2, s \right)
\Bigm \lvert_{\DY \;\equiv\; y_1 - y_2}
 \;,
\end{equation}
the `${\rm [accuracy]}$' superscript of ${\cal C}_0$ inclusively denoting $\NLLp$, $\HENLOp$, or $\LL$.
The $\delta (\DY - y_1 + y_2)$ function enforces the fixed-$\DY$ condition and thus removes one of the two rapidity integration: $y_2$ in our case.
The transverse momentum of the forward hadron (always a tetraquark) lie in the range $30 < |\bm{\kappa}_1| /{\rm GeV} < 120$,
whereas the one of the backward object (a singly heavy-flavored hadron or a light-flavored jet)
stays in the range $50 < |\bm{\kappa}_1| /{\rm GeV} < 120$.

These tailoring cuts allow the VFNS-based fragmentation approach to be valid, since energy scales are higher than thresholds for the DGLAP evolution of heavy quarks.
Furthermore, \emph{asymmetric} transverse-momentum windows permit to better disengage the pure resummation dynamics from the fixed-order background~\cite{Celiberto:2015yba,Celiberto:2015mpa,Celiberto:2020wpk}. 
They also suppress large Sudakov logarithms arising from (quasi) back-to-back emissions which are systematically missed by BFKL~\cite{Mueller:2013wwa,Marzani:2015oyb,Mueller:2015ael,Xiao:2018esv,Hatta:2020bgy,Hatta:2021jcd}.
Finally, they dampen instabilities encoded in higher-order contributions~\cite{Andersen:2001kta,Fontannaz:2001nq} and drastically reduce energy-momentum-conservation breaking effects~\cite{Ducloue:2014koa}.
Rapidity configurations are the typical ones of current LHC studies.
In particular,  hadrons are tagged only in the barrel calorimeter~\cite{Chatrchyan:2012xg}, say $|y_{1,2}| < 2.4$, while jets can be also detected by endcap detector~\cite{Khachatryan:2016udy}, say $|y_2| < 4.7$.

The second observable considered is the $|\bm{\kappa}_1|$-rate given by ${\cal C}_0$ coefficient, integrated over rapidities while $\DY$ is kept fixed, and integrated over $|\bm{\kappa}_2|$ but not over $|\bm{\kappa}_1|$
\begin{equation}
 \label{Yk1_rate}
 \frac{\drv \sigma(|\bm{\kappa}_1|, \DY, s)}{\drv |\bm{\kappa}_1| \drv \DY} =
 \int_{|\bm{\kappa}_1|^{\rm min}}^{|\bm{\kappa}_1|^{\rm max}} 
 \!\!\drv |\bm{\kappa}_1|
 \int_{\max \, (y_1^{\rm min}, \, y_2^{\rm min} + \DY)}^{\min \, (y_1^{\rm max}, \, y_2^{\rm max} + \DY)} \drv y_1
 \, \,
 {\cal C}_0^{\rm [accuracy]}\left(|\bm{\kappa}_1|, |\bm{\kappa}_2|, y_1, y_2, s \right)
\Bigm \lvert_{\DY \;\equiv\; y_1 - y_2}
 \;. 
\end{equation}

The last distribution considered is the $(|\bm{\kappa}_1| = |\bm{\kappa}_2|)$-rate, namely the ${\cal C}_0$ coefficient, integrated over rapidities while $\DY$ is kept fixed, differential in $|\bm{\kappa}_1|$ and $|\bm{\kappa}_2|$, but with the $|\bm{\kappa}_1| = |\bm{\kappa}_2|$ constraint enforced

\begin{equation}
 \label{Yk1eq2_rate}
 \frac{\drv \sigma(|\bm{\kappa}_1| = |\bm{\kappa}_2|, \DY, s)}{\drv |\bm{\kappa}_1| \drv |\bm{\kappa}_2| \drv \DY} =
 \int_{\max \, (y_1^{\rm min}, \, y_2^{\rm min} + \DY)}^{\min \, (y_1^{\rm max}, \, y_2^{\rm max} + \DY)} \drv y_1
 \, \,
 {\cal C}_0^{\rm [accuracy]}\left(|\bm{\kappa}_1|, |\bm{\kappa}_2| \equiv |\bm{\kappa}_1|, y_1, y_2, s \right)
\Bigm \lvert_{\DY \;\equiv\; y_1 - y_2}
 \;. 
\end{equation}

Both the $|\bm{\kappa}_1|$- and the $(|\bm{\kappa}_1| = |\bm{\kappa}_2|)$-rate rapidities stay in the same range which the $\DY$-distribution it tailored on, while the nonintegrated transverse momenta span from 10~to~120~GeV.
To study configurations that align well with prospective HL-LHC data, we average our predictions on  transverse-momentum bins set at a constant width of 10~GeV.
Then, $\DY$ will be integrated in a characteristic forward bin, say $3 < \DY < 4.5$ for the tetraquark-plus-hadron case, and $4 < \DY < 6$ for the tetraquark-plus-jet one.

\subsection{Rapidity-interval rates}
\label{ssec:Y_rates}

\begin{figure*}[!t]
\centering

   \includegraphics[scale=0.41,clip]{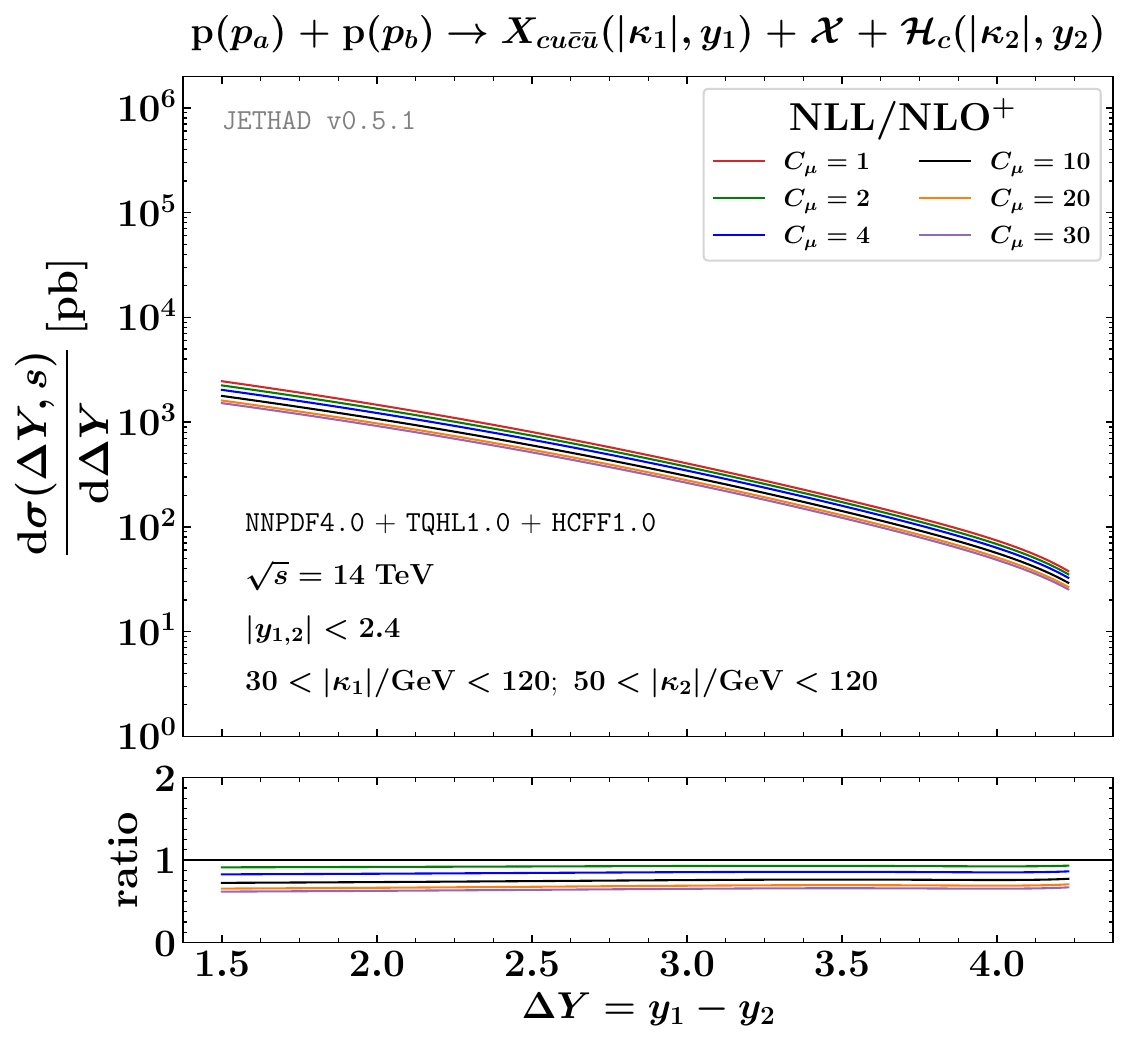}
   \hspace{0.15cm}
   \includegraphics[scale=0.41,clip]{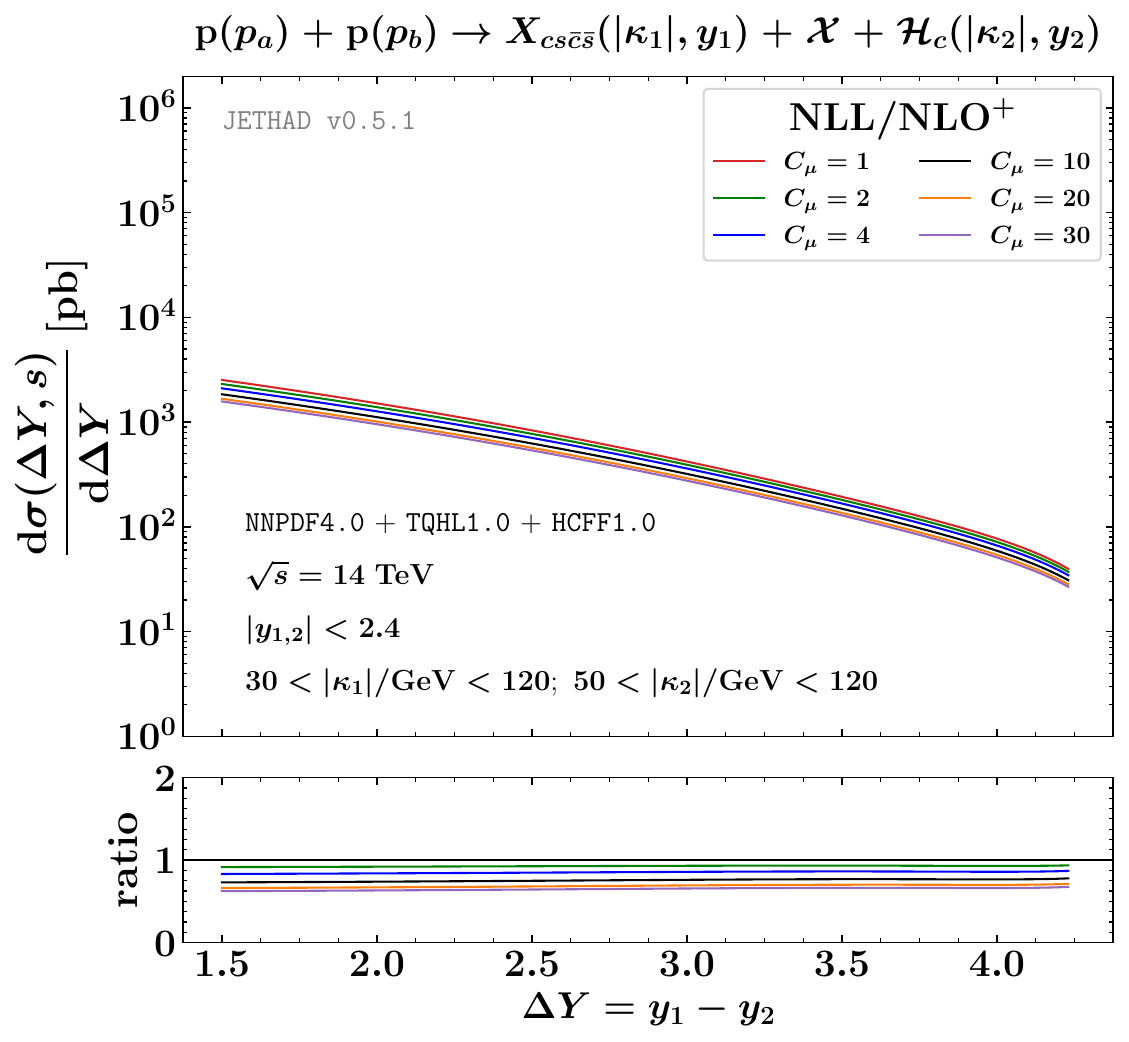}

   \vspace{0.75cm}

   \includegraphics[scale=0.41,clip]{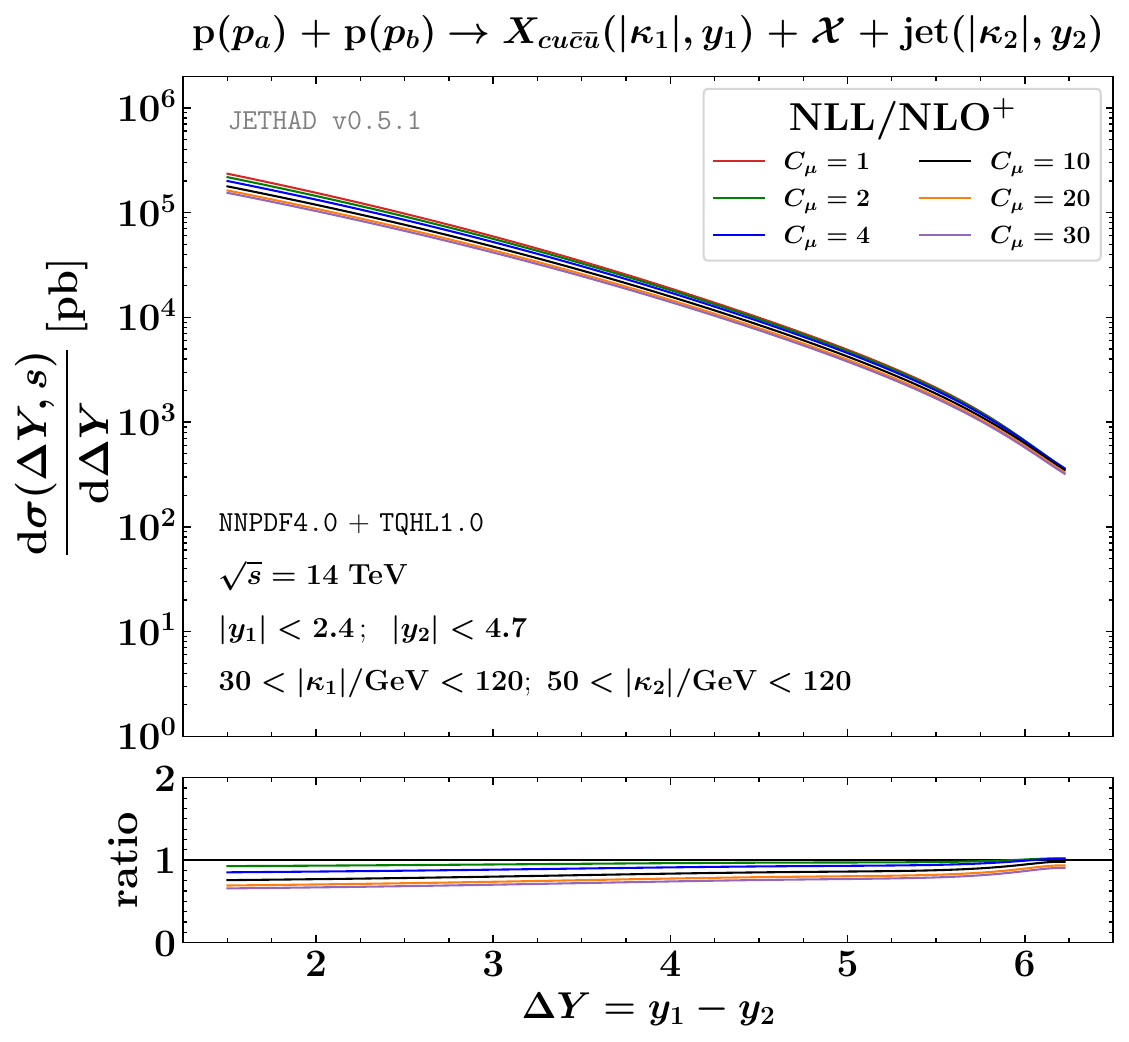}
   \hspace{0.15cm}
   \includegraphics[scale=0.41,clip]{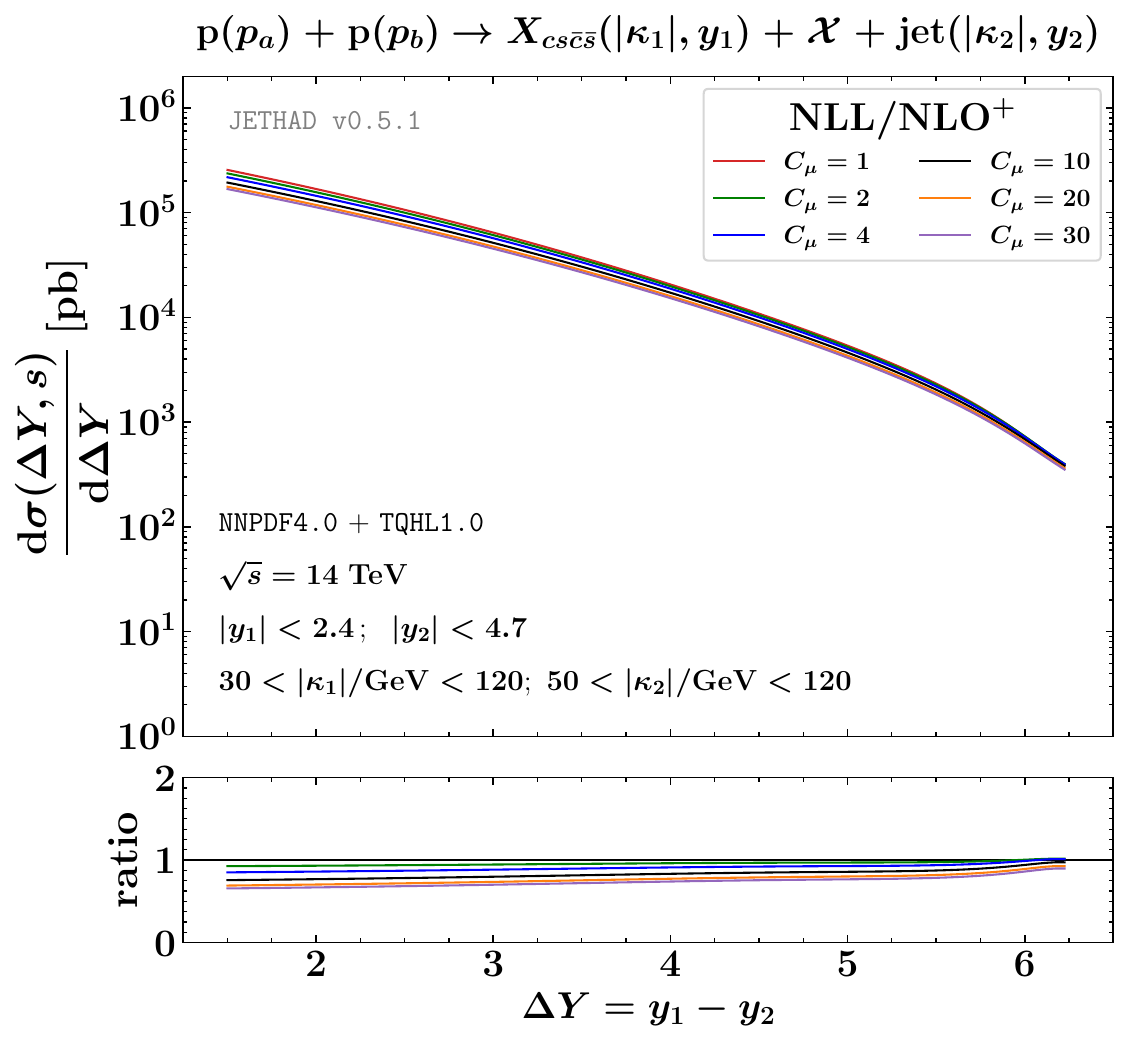}

\caption{$\NLLp$ versus $\HENLOp$ $\DY$-rates for $\Xcq + {\cal H}_Q$ (upper) and $\Xcq + {\rm jet}$ (lower) reactions at 14~TeV~LHC. An extended study of MHOUs in the range $1 < C_\mu < 30$ is illustrated.}
\label{fig:Y_cq_psv}
\end{figure*}

\begin{figure*}[!t]
\centering

   \includegraphics[scale=0.41,clip]{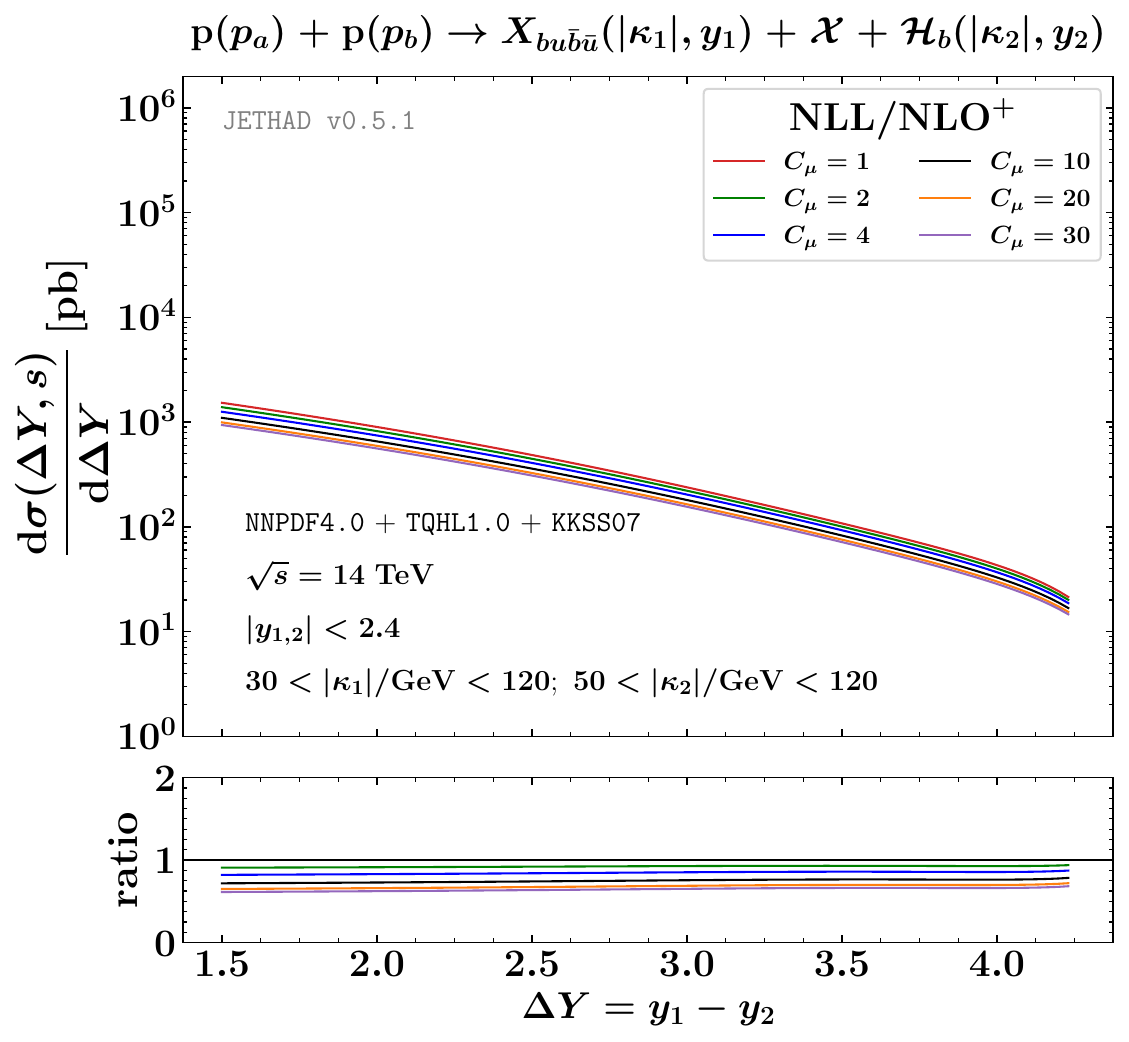}
   \hspace{0.15cm}
   \includegraphics[scale=0.41,clip]{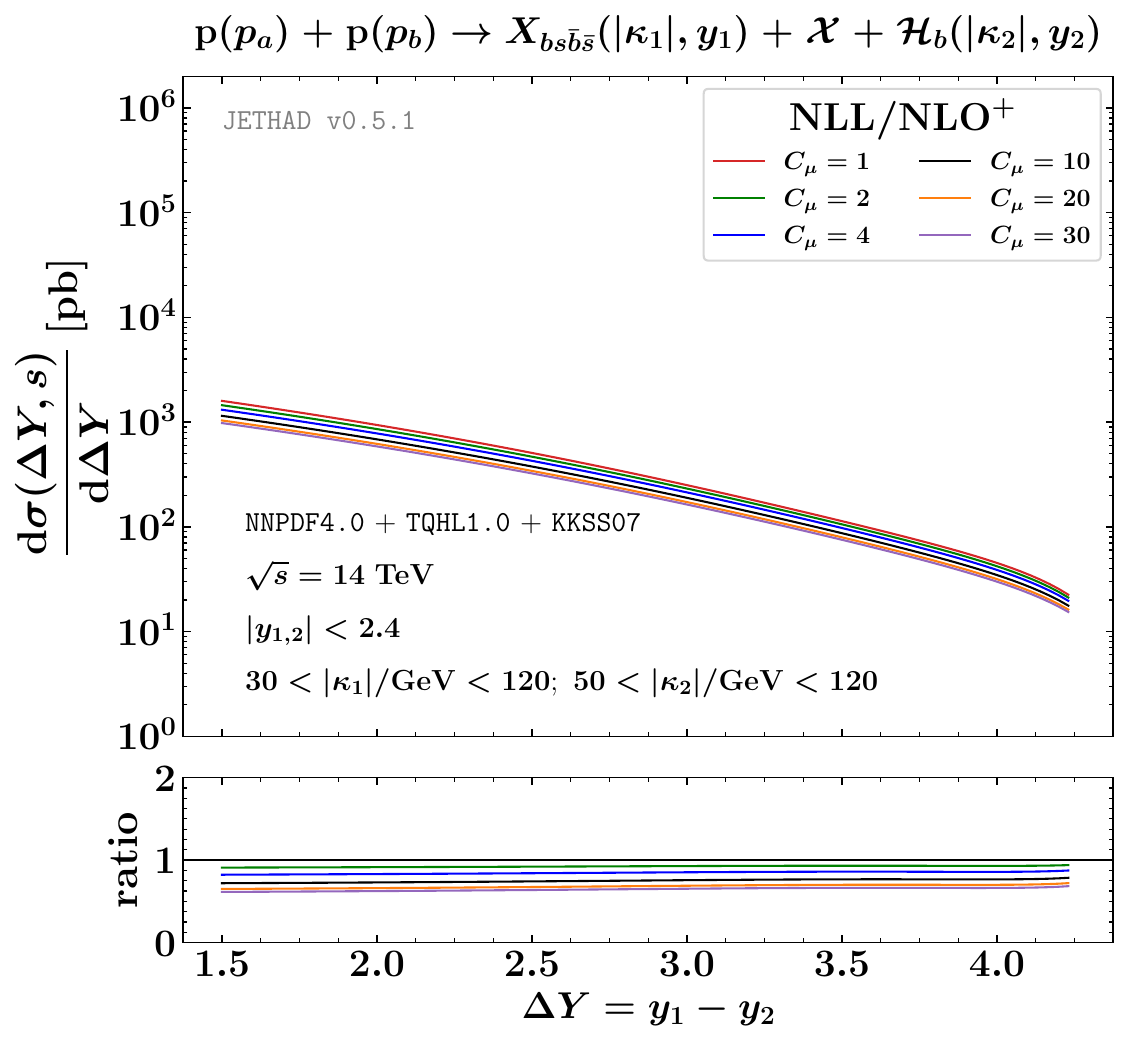}

   \vspace{0.75cm}

   \includegraphics[scale=0.41,clip]{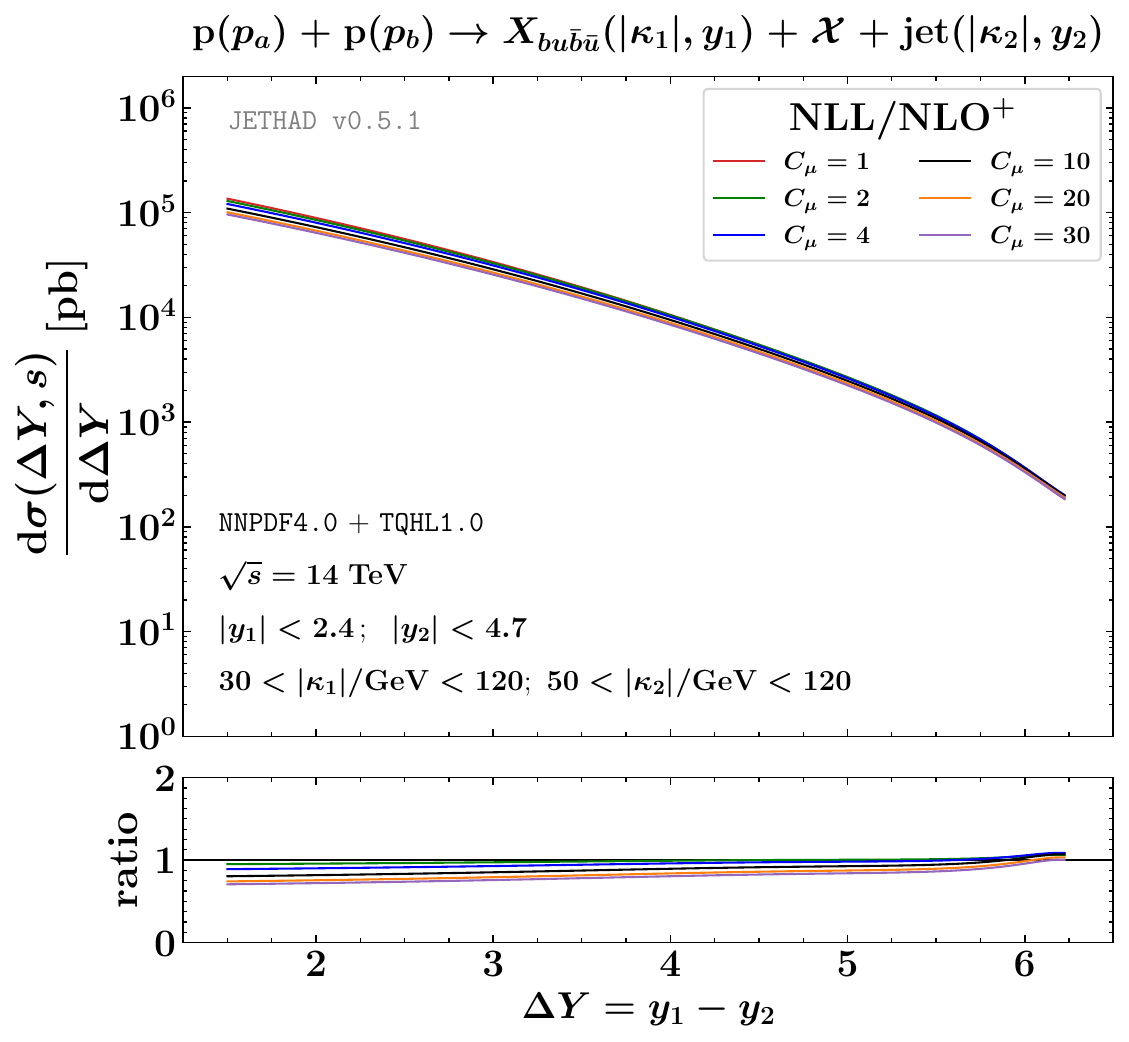}
   \hspace{0.15cm}
   \includegraphics[scale=0.41,clip]{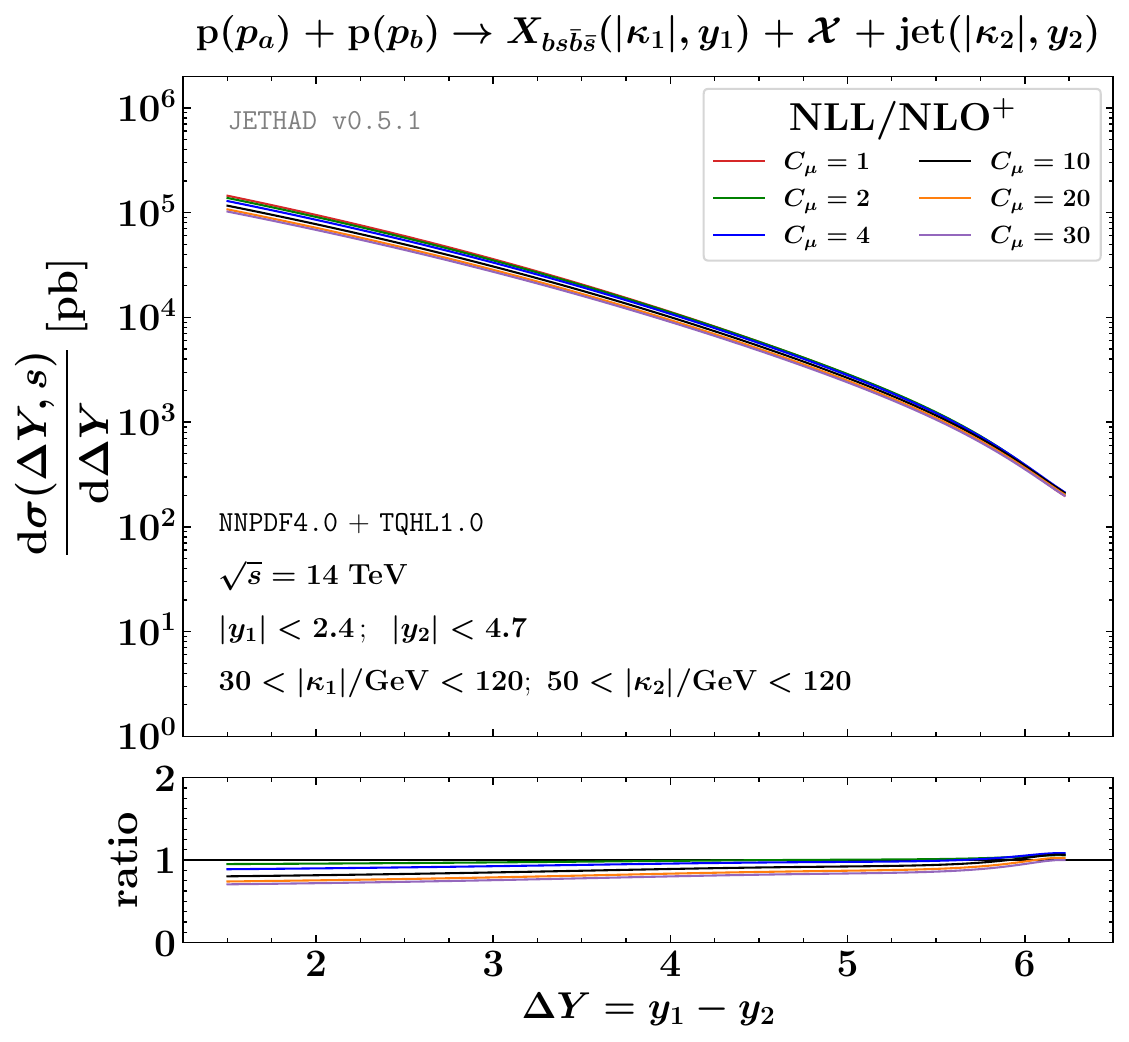}

\caption{$\NLLp$ versus $\HENLOp$ $\DY$-rates for $\Xbq + {\cal H}_Q$ (upper) and $\Xbq + {\rm jet}$ (lower) reactions at 14~TeV~LHC. An extended study of MHOUs in the range $1 < C_\mu < 30$ is illustrated.}
\label{fig:Y_bq_psv}
\end{figure*}

\begin{figure*}[!t]
\centering

   \includegraphics[scale=0.41,clip]{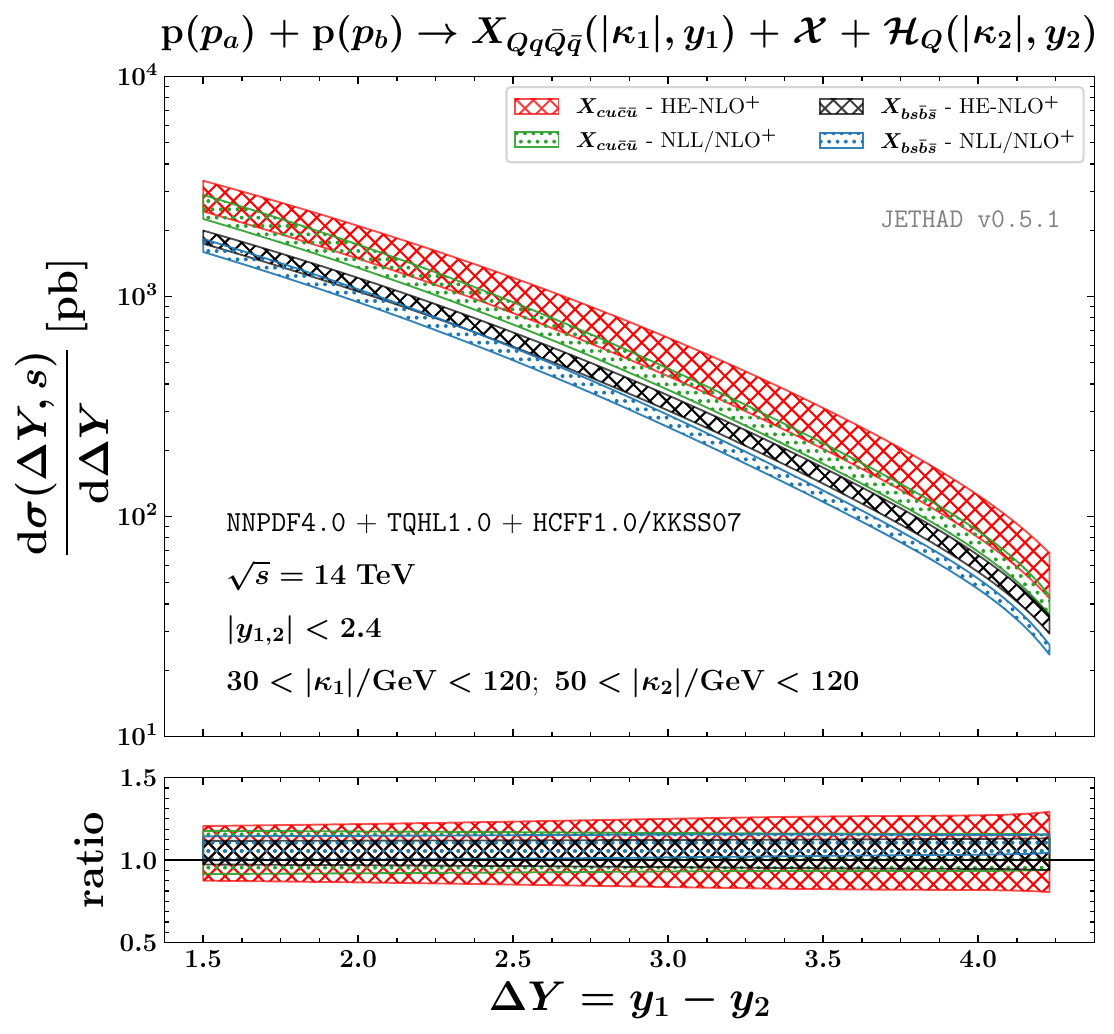}
   \hspace{0.15cm}
   \includegraphics[scale=0.41,clip]{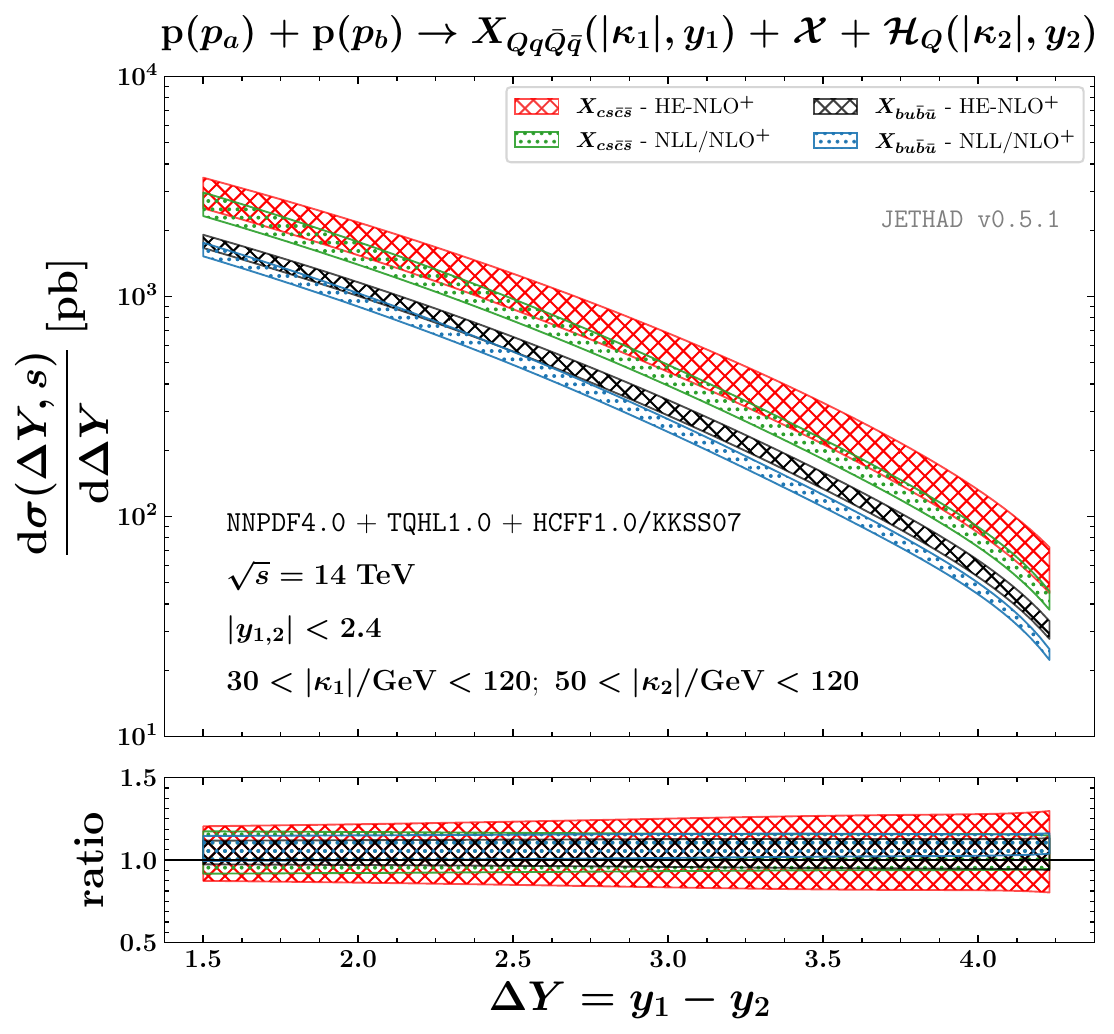}

   \vspace{0.75cm}

   \includegraphics[scale=0.41,clip]{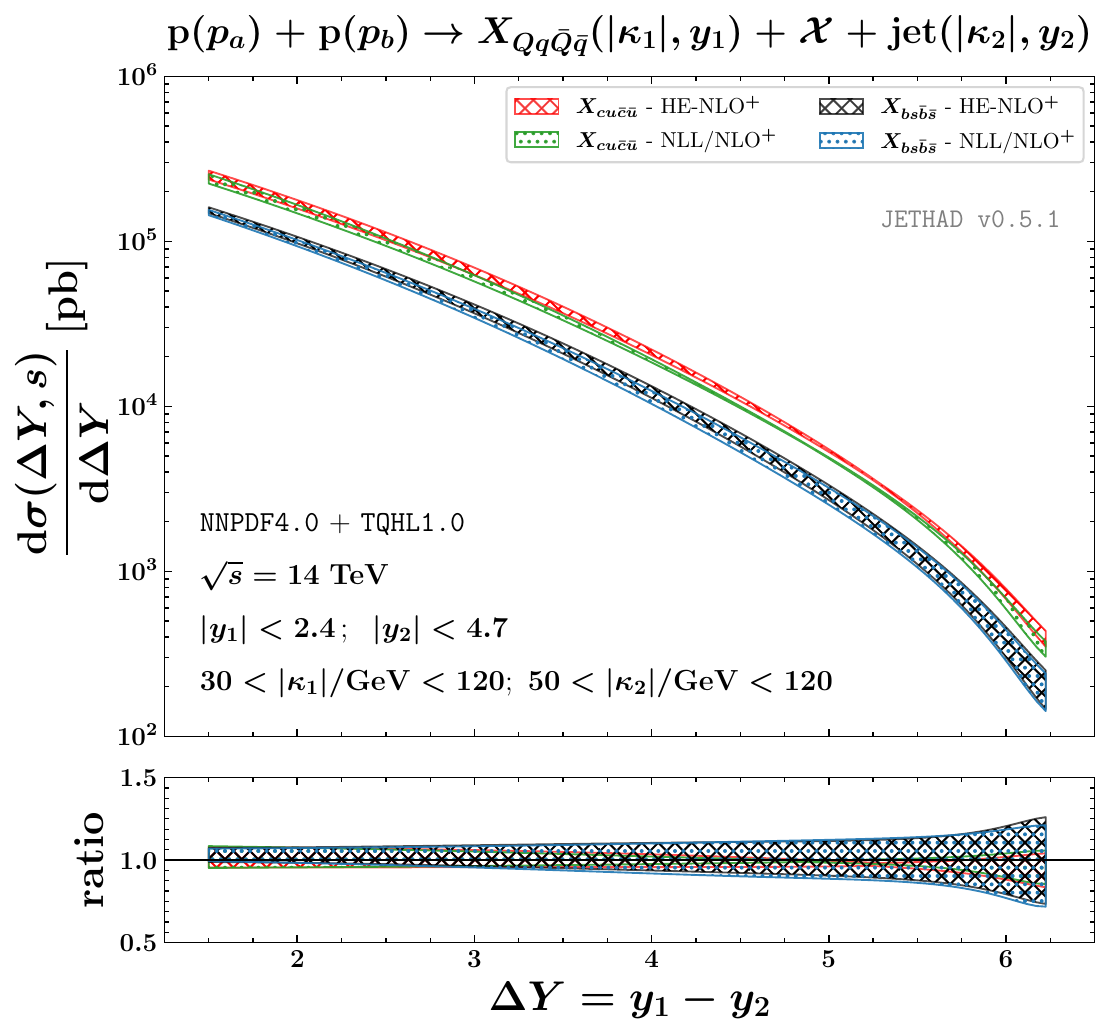}
   \hspace{0.15cm}
   \includegraphics[scale=0.41,clip]{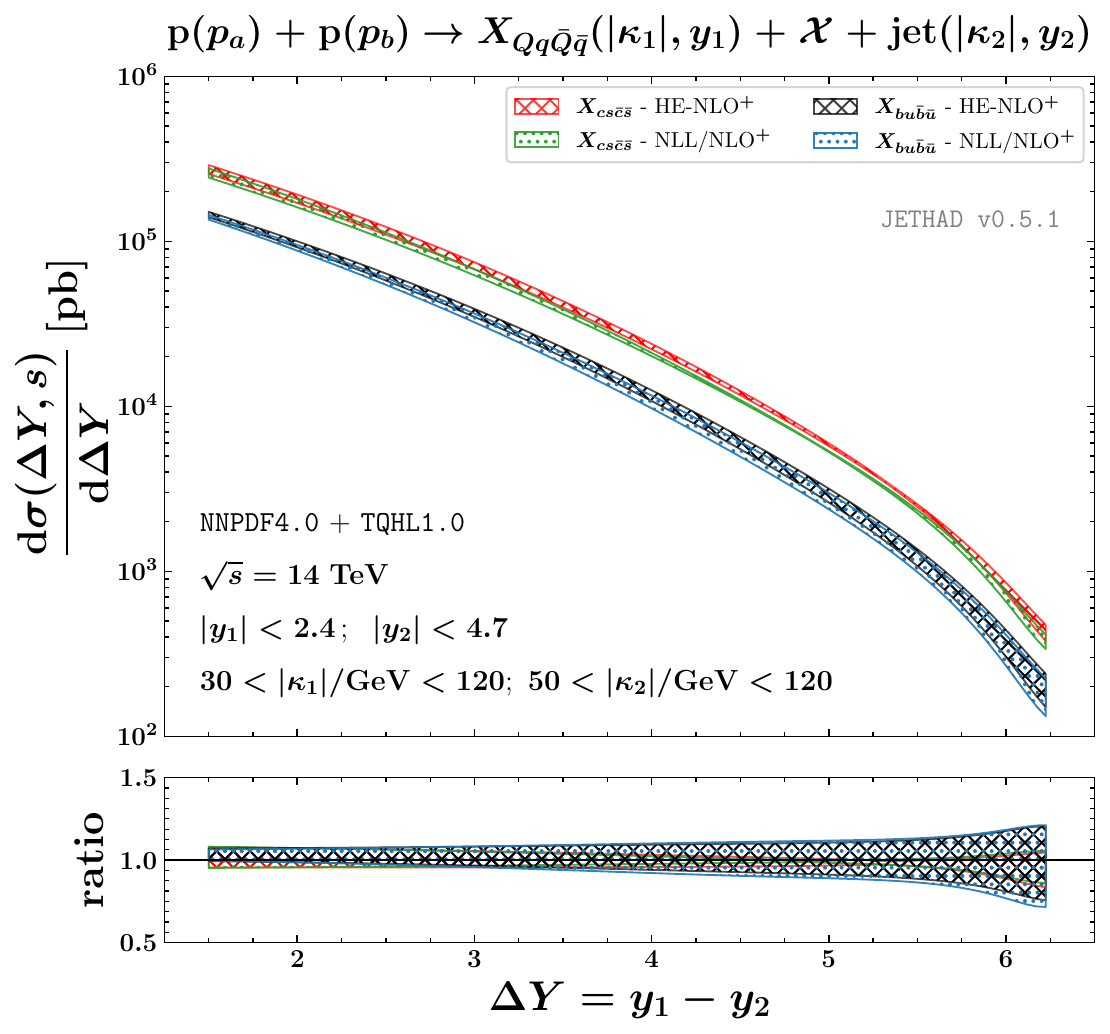}

\caption{$\NLLp$ versus $\HENLOp$ $\DY$-rates for $\XQq + {\cal H}_Q$ (upper) and $\XQq + {\rm jet}$ (lower) reactions at 14~TeV~LHC. Uncertainty bands are for the combined effect of MHOUs and errors on numeric integrations.}
\label{fig:Y_MHOU}
\end{figure*}

In this Section we present predictions for the rapidity-interval distribution for our reference processes (see Fig.~\ref{fig:process}).
The main analysis depicted in Figs.~\ref{fig:Y_cq_psv} and~\ref{fig:Y_bq_psv} entails an examination of the observable across a wide spectrum.
It is achieved through a progressive variation of factorization and renormalization scales, which spans a broad window, regulated by the parameter $C_\mu$ introduced in Section~\ref{ssec:uncertainty} and ranging from $1$ to $30$.
This approach expands upon the conventional MHOU scan, which typically operates within the range $1/2 < C_\mu < 2$.

Left (right) plots of Fig.~\ref{fig:Y_cq_psv} contain $\NLLp$ results for $\Xcu$ ($\Xcs$) production channels, whereas left (right) plots of Fig.~\ref{fig:Y_bq_psv} embody predictions for $\Xbu$ ($\Xbs$) corresponding ones.
In all cases, upper (lower) plots are for tetraquark-plus-hadron (tetraquark-plus-jet) reactions.
Ancillary panels below the primary ones display normalized distributions, obtained by dividing each distribution by its central value computed at $C_\mu = 1$.

The general outcome from all eight examined final states is a minimal sensitivity to variations in $C_\mu$ across the entire range of $\DY$ values explored in our analysis. This observation underscores the significant stabilizing mechanism embedded within our {\tt TQHL1.0} functions.

Results of Fig.~\ref{fig:Y_MHOU} are for rapidity-interval distributions with a standard MHOU analysis in the range range $1/2 < C_\mu < 2$.
The overall decreasing pattern with $\DY$ of our $\DY$-rates results from the interplay of two contrasting tendencies: while BFKL partonic cross sections increase with $\DY$ and subsequently with energy, their convolution with collinear PDFs and FFs in the impact factors markedly restrains this escalation.
Ancillary panels below primary plots in Fig.~\ref{fig:Y_MHOU} underscore the stabilizing power of our {\tt TQHL1.0} FFs, with the $\NLLp$ uncertainty bands consistently staying within the $\LL$ ones for all channels.

We mention a pertinent aspect that extends the topic beyond the scope of this review, but still deserves a discussion. We refer to the influence of \ac{MPIs} on differential distributions as the final-state rapidity distance $\DY$ increase.
Particularly noteworthy is the effect of the \ac{DPS}. 
A recent analysis quantifies DPS corrections to Mueller--Navelet jets, potentially impacting $\DY$-distributions at high center-of-mass energies and moderate transverse momenta~\cite{Ducloue:2015jba}.

Concerning quarkonia, studies of final states such as double $\Jpsi$~\cite{Lansberg:2014swa,Lansberg:2020rft}, $\Jpsi$~plus~$\Yps$~\cite{Lansberg:2020rft}, $\Jpsi$ plus electroweak-boson~\cite{Lansberg:2016rcx,Lansberg:2017chq}, and triple $\Jpsi$~\cite{dEnterria:2016ids,Shao:2019qob}, have indicated the presence potentially large DPS contributions.
Thus, the search for MPI signatures also in tetraquark-sensitive final states represents an important and promising avenue for future research.

\subsection{Transverse-momentum rates}
\label{ssec:pT_rates}

\begin{figure*}[!t]
\centering

   \includegraphics[scale=0.41,clip]{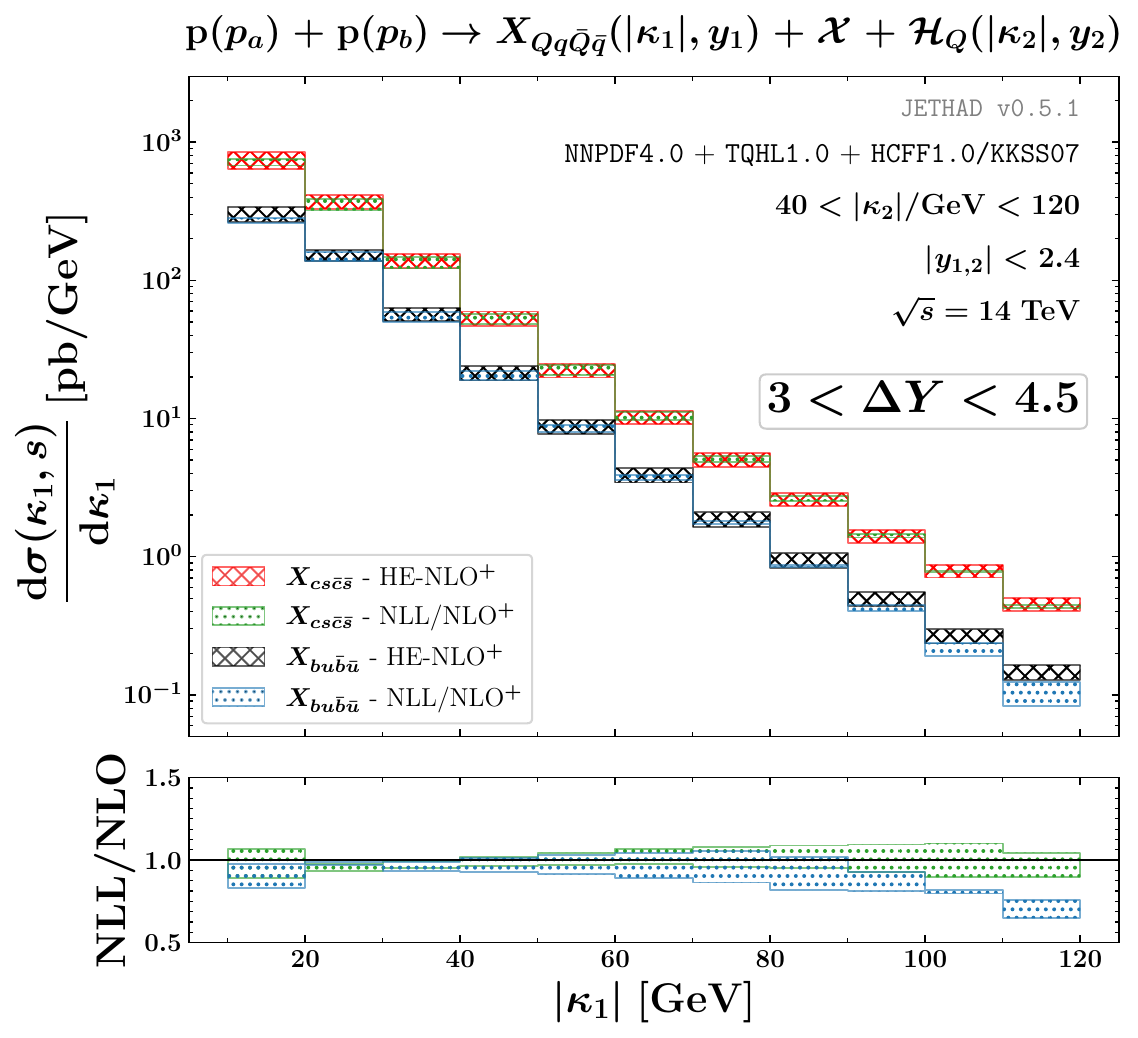}
   \hspace{0.15cm}
   \includegraphics[scale=0.41,clip]{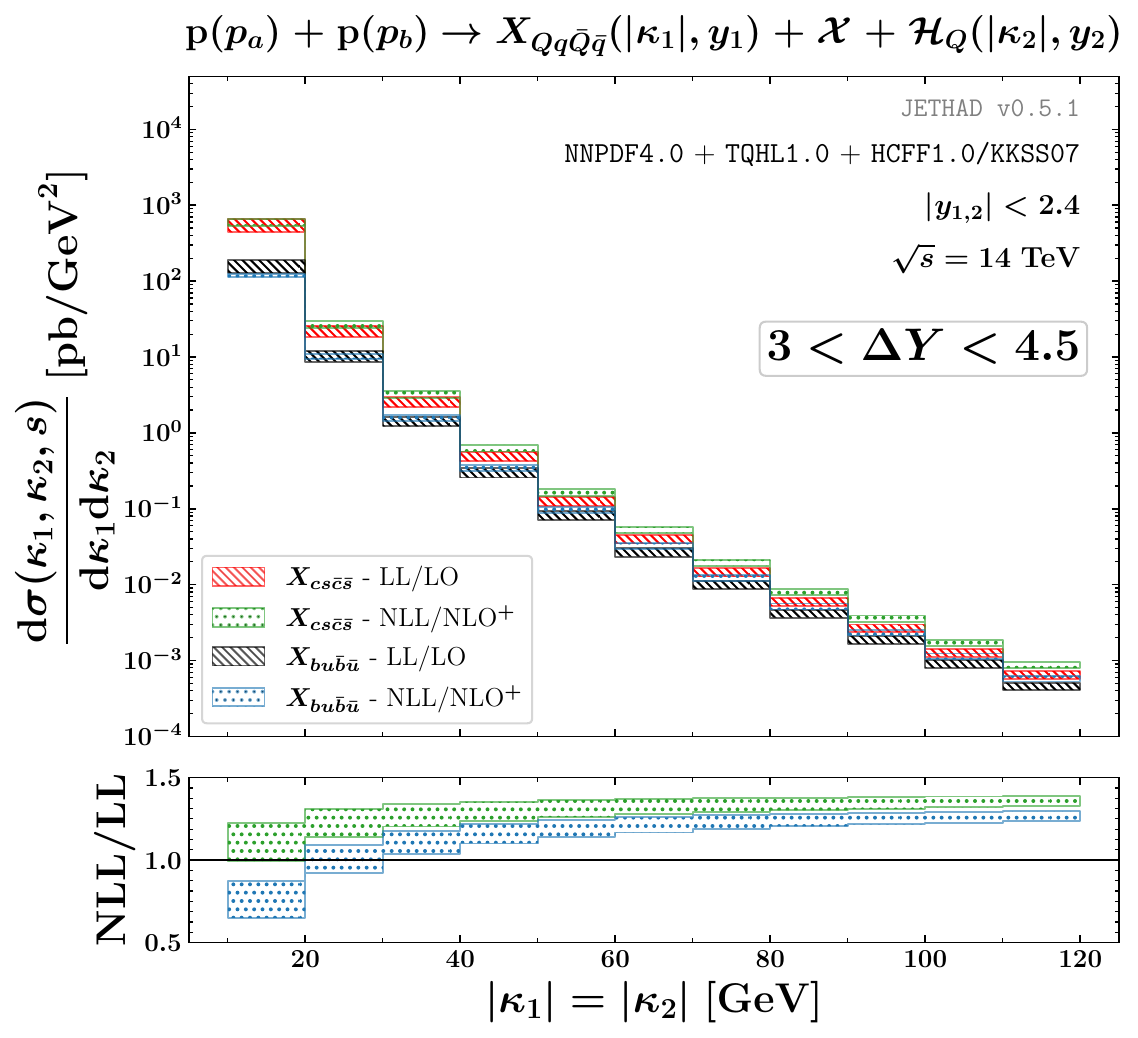}

   \vspace{0.75cm}

   \includegraphics[scale=0.41,clip]{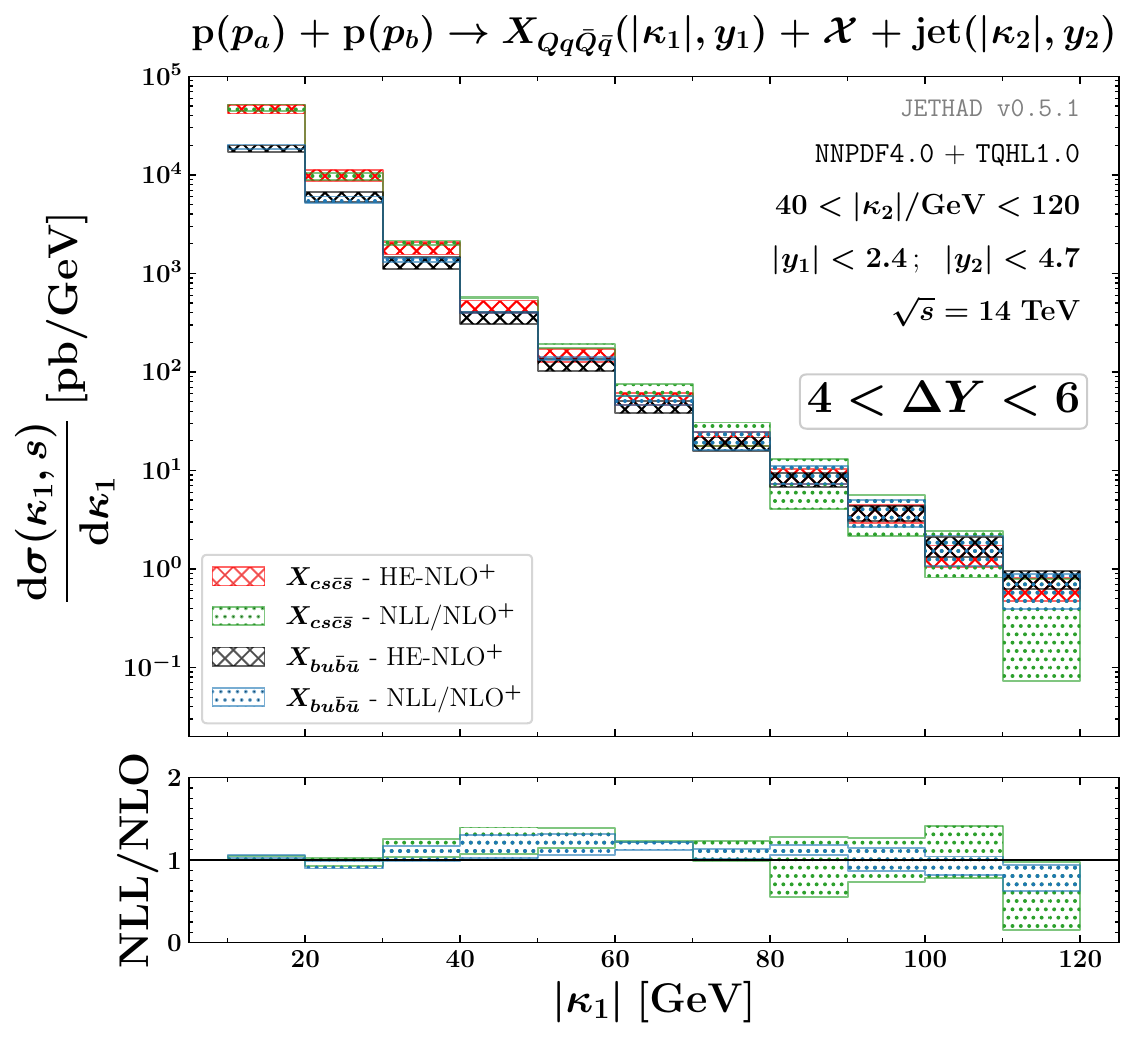}
   \hspace{0.15cm}
   \includegraphics[scale=0.41,clip]{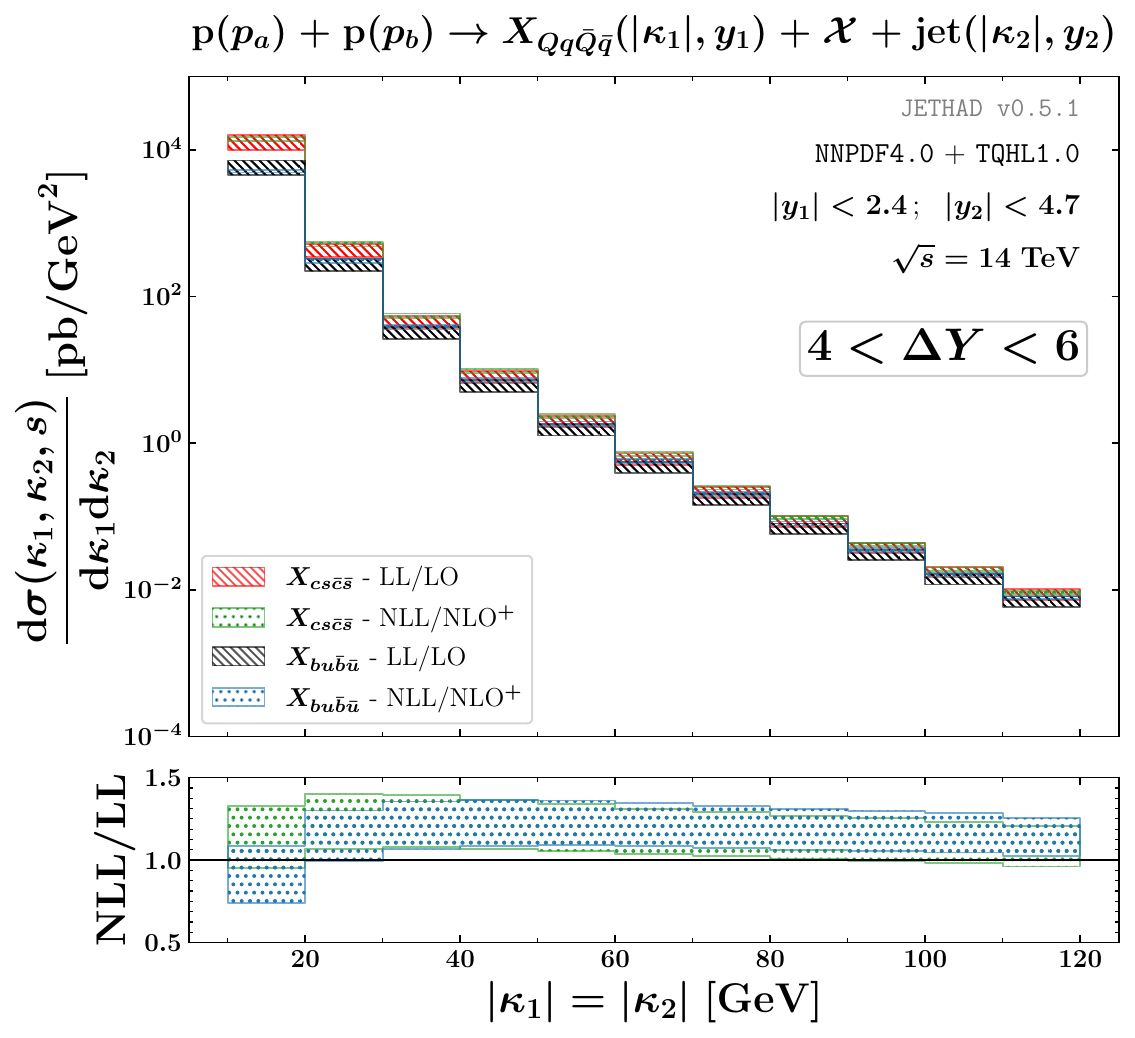}

\caption{$\NLLp$ transverse-momentum rates for $\XQq + {\cal H}_Q$ (upper) and $\XQq + {\rm jet}$ (lower) reactions at 14~TeV~LHC. Uncertainty bands are for the combined effect of MHOUs and errors on numeric integrations.}
\label{fig:pT_MHOU}
\end{figure*}

The first transverse-momentum distribution matter of our investigation is the $|\bm{\kappa}_1|$-rate.
This observable serves as a common basis for exploring potential links between our hybrid factorization and alternative formalisms. By varying $|\vec \kappa_1|$ within a wide range, we can delve into a broad kinematic domain, where alternative resummation techniques may become pertinent and necessary.

In scenarios where transverse momenta are high or widely separated, the magnitude of DGLAP-like logarithms and threshold contaminations increases, rendering a pure high-energy approach inadequate~\cite{Sterman:1986aj,Catani:1989ne,Catani:1996yz,Bonciani:2003nt,deFlorian:2005fzc,Ahrens:2009cxz,deFlorian:2012yg,Forte:2021wxe,Mukherjee:2006uu,Bolzoni:2006ky,Becher:2006nr,Becher:2007ty,Bonvini:2010tp,Ahmed:2014era,Banerjee:2018vvb,Duhr:2022cob,Shi:2021hwx,Wang:2022zdu}. 
Additionally, in the very-low transverse-momentum regime, enhanced $|\vec \kappa_1|$-logarithms, which are not accounted for by BFKL, become significant. 
Moreover, effects like diffusion patterns grow to a point where they impede the convergence of high-energy resummation~\cite{Bartels:1993du,Caporale:2013bva,Ross:2016zwl}.

In these kinematic corners, an all-order transverse-momentum (TM) resummation is necessary~\cite{Catani:2000vq,Bozzi:2005wk,Bozzi:2008bb,Catani:2010pd,Catani:2011kr,Catani:2013tia,Catani:2015vma,Duhr:2022yyp}. 
TM-resummed distributions have been explored extensively in various processes, including photon~\cite{Cieri:2015rqa,Alioli:2020qrd,Becher:2020ugp,Neumann:2021zkb}, Higgs~\cite{Ferrera:2016prr}, and $W$-boson pair production, as well as in final states involving bosons and jets.
Recent studies have provided third-order fiducial predictions for Drell--Yan and Higgs emissions, incorporating TM resummation~\cite{Ebert:2020dfc,Re:2021con,Chen:2022cgv,Neumann:2022lft,Bizon:2017rah,Billis:2021ecs,Re:2021con,Caola:2022ayt}.
Furthermore, when the transverse momenta of detected particles result in nearly back-to-back configurations, Sudakov-type logarithms emerge, requiring additional resummation techniques~\cite{Mueller:2012uf,Mueller:2013wwa,Marzani:2015oyb,Mueller:2015ael,Xiao:2018esv}.

Left panels of Fig.~\ref{fig:pT_MHOU} carry information about $|\bm{\kappa}_1|$-rates for tetraquark-plus-hadron (upper) and tetraquark-plus-jet (right) channels with their within $\NLLp$ and $\HENLOp$ accuracies.
Here, for the sake of brevity, we consider just the $\Xcs$ and $\Xbu$ channels, namely the ones complementary to our first study presented in Ref.~\cite{Celiberto:2023rzw}.
Overall, we observe a falloff with increasing $|\bm{\kappa}_1|$ across the distributions. Notably, the results remain remarkably stable MHOU studies, with error bands generally confined within a 30\% width, except for the initial and final two bins.

In the tetraquark-plus-hadron scenario, the NLL-resummed distribution initially appears smaller than its fixed-order counterpart in the first bin, subsequently reaching a comparable magnitude and exhibiting a slight upward trend with $|\bm{\kappa}_1|$. 
Conversely, in the case of hadron-plus-jet events, the resummation leads to a visible rise with $|\bm{\kappa}_1|$, reaching up to a 50\% increase. 
However, larger uncertainties in the final two bins indicate a loss of stability for BFKL due to threshold contaminations. 
Notably, these uncertainties are less pronounced when detecting a bottomed tetraquark, aligning with recent findings suggesting that VFNS FFs for bottom-flavored hadrons offer greater stabilizing effects compared to their charm-flavored counterparts~\cite{Celiberto:2021fdp}.

On the other hand, the initial bin may be susceptible to instabilities arising from energy scales nearing thresholds for DGLAP evolution dictated by heavy-quark masses. This could explain the downturn observed in the $\Xbu$ NLL cross section with respect to the $\Xcs$ one in the left upper plot, which could potentially stem from instabilities associated with values of $\mu_F$ approaching the bottom mass.

The final observable under investigation is a doubly differential distribution in both $|\bm{\kappa}_1|$ and $|\bm{\kappa}_2|$.
As the separation between the two transverse momenta increases, additional kinematic regions adjacent to the BFKL regime become accessible (as observed in recent analyses on ${\cal H}_b$ hadrons~\cite{Celiberto:2021fdp} and $\Xi$ baryons~\cite{Celiberto:2022kxx}).
A joint resummation of transverse-momentum logarithms for two-particle distributions was initially achieved in the context of Higgs-plus-jet hadroproduction~\cite{Monni:2019yyr} through the {\RadISH} momentum-space method~\cite{Bizon:2017rah}.
Here we make use of a complementary configuration by setting $|\bm{\kappa}_1| \equiv |\bm{\kappa}_2|$ and spanning them from 10~to~120~GeV, while maintaining the same rapidity bins as before. This choice permits to examine in deep the strict BFKL regime, thus easing a precise determination of the impact of higher-order corrections.

In the right plots of Fig.\tref{fig:pT_MHOU} we present $\NLLp$ distributions for both the tetraquark-plus-hadron (upper) and hadron-plus-jet (lower) channel, juxtaposed with their $\LL$ limits. To provide further insight, we magnify the NLL to LL ratio in ancillary plots. With the exception of the initial bin, where $\Xbu$ cross sections are affected by instabilities due to their proximity to thresholds, NLL corrections in the upper plot exhibit a moderate increase with transverse momentum, plateauing at around +30\%. Conversely, NLL corrections in the lower plot consistently augment LL results by approximately +30\% across the entire spectrum. This dichotomy arises from the positive sign of NLO hadron impact-factor corrections owing to large values of the $(gg)$ channel~\cite{Celiberto:2017ptm}, while NLO jet corrections are generally negative~\cite{Ducloue:2013bva,Caporale:2014gpa,Celiberto:2015yba,Celiberto:2022gji}.

\section{Final remarks}
\label{sec:conclusions}

We made use of the hybrid high-energy and collinear factorization framework at $\NLLp$ to conduct a comprehensive study of the inclusive hadroproduction of four different species of exotic heavy-light tetraquark at high energies.
We described the formation of tetraquark states by means of the single-parton collinear-fragmentation mechanism at leading twist, which holds validity in the large transverse-momentum regime pertinent to our analysis.

To this end, we developed first tetraquark collinear NLO FF sets, named {\tt TQHL1.0} functions. 
They were constructed by DGLAP-evolving a SNAJ-inspired model input for the heavy-quark channel~\cite{Suzuki:1985up,Nejad:2021mmp,Suzuki:1977km,Amiri:1986zv}.
The interconnection between high-energy resummation and the fragmentation mechanism played a crucial role in our analysis.

Notably, we observed that the distinct behavior of the gluon-to-tetraquark fragmentation channel acts as a robust stabilizer of the hybrid factorization scheme under NLL corrections and scale variations. This enabled us to achieve a remarkable level of accuracy in describing our observables at the natural scales provided by kinematics.
Natural stability, revealed as a general property across various heavy-flavored species considered thus far, including single heavy-flavored hadrons~\cite{Celiberto:2021dzy,Celiberto:2021fdp,Celiberto:2022rfj,Celiberto:2022zdg}, vector quarkonia~\cite{Celiberto:2022dyf,Celiberto:2023fzz}, and charmed $B$ mesons~\cite{Celiberto:2022keu,Celiberto:2024omj}), was now found also in the case of $\XQq$ states as well.

As for future tetraquark studies within high-energy QCD, a step forward would come out from investigations on single-forward tetraquark detections within our hybrid-factorization formalism.
These channels would give us a direct access to the small-$x$ proton UGD, whose current knowledge is very qualitative and is mainly based on models.

Once a more precise description of the UGD becomes available, it will enable a direct comparison with upcoming data collected at the HL-LHC.
We mainly refer to semi-inclusive final states sensitive to the single-inclusive production of tetraquark states in central rapidity regions covered by ATLAS or CMS barrels, say $|y_1| \lesssim 2.4$, or in forward regions at LHCb, say $2 \lesssim y_1 \lesssim 4.5$.
Accessing a wider range of transverse momentum will also permit to probe kinematic sectors where different formation mechanisms for our tetraquark are supposed to be at work, ad possibly establish a data-driven hierarchy among them.

Further advancements would also rely upon linking our program with NLO investigations of single-forward or nearly back-to-back semi-inclusive emissions within the framework of gluon saturation (see, for instance, Refs.~\cite{Gelis:2010nm,Kovchegov:2012mbw,Chirilli:2012jd,Boussarie:2014lxa,Benic:2016uku,Benic:2018hvb,Roy:2019hwr,Roy:2019cux,Beuf:2020dxl,Iancu:2021rup,Iancu:2023lel,vanHameren:2023oiq,Wallon:2023asa} and references therein).
In these analyses, the influence of soft-gluon radiation on angular asymmetries in dijet or dihadron production was addressed (see Refs.~\cite{Hatta:2020bgy,Hatta:2021jcd,Caucal:2021ent,Caucal:2022ulg,Taels:2022tza,Fucilla:2022wcg}).

NLO saturation allows us to access the (un)polarized gluon content of protons and nucleons at small-$x$~\cite{Kotko:2015ura,vanHameren:2016ftb,Altinoluk:2020qet,Altinoluk:2021ygv,Boussarie:2021ybe,Caucal:2023nci}.
Authors of Refs.~\cite{Kang:2013hta,Ma:2014mri,Ma:2015sia,Ma:2018qvc,Stebel:2021bbn} explore heavy-hadron emissions in proton-proton and proton-nucleus collisions by incorporating small-$x$ effects.
Future investigations will explore the intersection between our approach to tetraquarks production via the $\NLLp$ hybrid factorization and higher-order calculations for NLO saturation in exclusive emissions of heavy particles~\cite{Mantysaari:2021ryb,Mantysaari:2022kdm}.

A milestone in deepening our understanding of tetraquark fragmentation will rely upon comparing predictions from our {\tt TQHL1.0} FFs with ones based on functions extracted from global data.
Along this direction, artificial-intelligence techniques already adapted to address the collinear fragmentation of lighter-hadron species~\cite{Nocera:2017qgb,Bertone:2017xsf,Bertone:2017tyb,Bertone:2018ecm,Khalek:2021gxf,Khalek:2022vgy,Soleymaninia:2022qjf,Soleymaninia:2022alt} will be an asset.

As a prospect, we plan to extend to explore heavy-light tetraquark production through other resummations, as well as by using different inputs for the initial-scale fragmentation and through exclusive reactions. Promising avenues to be investigated via the fragmentation approximation include emissions of fully-charmed tetraquarks~\cite{Feng:2020riv,Feng:2020qee,Feng:2023agq,Feng:2023ghc} and pentaquarks~\cite{Cheung:2004ji,MoosaviNejad:2023qew,Farashaeian:2024son}.

A window of novel opportunities will come by forthcoming advancements on $X(3872)$ spectroscopy at electron-hadron facilities~\cite{Albaladejo:2020tzt,Winney:2022tky,Winney:2024hhs}.
Until recently, the $X(3872)$ was the sole exotic state observed in prompt proton collisions. However, the discovery of fully-charmed structures~\cite{LHCb:2020bwg} and the $T_{cc}$~\cite{LHCb:2021vvq,LHCb:2021auc} brought novelty to the exotic-physics panorama. 
Extending our fragmentation approach to these states should be feasible and could offer valuable insights. Additionally, it could be applied to investigate the $Z_c(3900)$, which has not yet been observed promptly~\cite{Guo:2013ufa}.

While further analyses are required from both the theoretical and phenomenological perspectives, we believe that the present study can contribute to open new windows of discovery. This gives us an intriguing opportunity to gain deeper insights into the true nature of exotic matter, which can be explored at the HL-LHC and at future colliders~\cite{Chapon:2020heu,Anchordoqui:2021ghd,Feng:2022inv,Hentschinski:2022xnd,Accardi:2012qut,AbdulKhalek:2021gbh,Khalek:2022bzd,Acosta:2022ejc,AlexanderAryshev:2022pkx,Brunner:2022usy,Arbuzov:2020cqg,Abazov:2021hku,Bernardi:2022hny,Amoroso:2022eow,Celiberto:2018hdy,Klein:2020nvu,2064676,MuonCollider:2022xlm,Aime:2022flm,MuonCollider:2022ded,Accettura:2023ked,Vignaroli:2023rxr,Black:2022cth,Dawson:2022zbb,Bose:2022obr,Begel:2022kwp,Abir:2023fpo,Accardi:2023chb}.

\section*{Data availability}
\label{sec:data_availability}
\addcontentsline{toc}{section}{\nameref{sec:data_availability}}

The four grids for the {\tt TQHL1.0} FF determinations
\begin{itemize}
    \item NLO, $\Xbu$\,: \,{\tt TQHL10\_Xbu\_nlo};
    \item NLO, $\Xcu$\,: \,{\tt TQHL10\_Xcu\_nlo};
    \item NLO, $\Xbs$\,: \,{\tt TQHL10\_Xbs\_nlo};
    \item NLO, $\Xcs$\,: \,{\tt TQHL10\_Xcs\_nlo},
\end{itemize}
can be publicly accessed from the following url: \url{https://github.com/FGCeliberto/Collinear_FFs/}.

Data underlying figures presented in this review can be made available upon a reasonable request.

\section*{Acknowledgments}
\label{sec:acknowledgments}
\addcontentsline{toc}{section}{\nameref{sec:acknowledgments}}

The author thanks colleagues of \textbf{Quarkonia As Tools} and \textbf{EXOTICO} conferences for inspiring discussions and the welcoming atmosphere.
The author is grateful to Alessandro~Papa, Seyed~Mohammad~Moosavi~Nejad, Annalisa~D'Angelo, Alessandro~Pilloni, and Ingo~Schienbein for fruiful conversations.
This work was supported by the Atracci\'on de Talento Grant n. 2022-T1/TIC-24176 of the Comunidad Aut\'onoma de Madrid, Spain.

\begin{appendices}

\printacronyms

\setcounter{appcnt}{0}
\hypertarget{app:NLOHEF}{
\section*{Appendix~A: NLO correction for the heavy-hadron singly off-shell emission function}}
\label{app:NLOHEF}

The analytic expression for the NLO correction to the forward heavy-hadron singly off-shell emission function reads~\cite{Ivanov:2012iv}

\begin{equation}
  \label{NLOHEF}
  \hat \F_h(n,\nu,|\bm{\kappa}_h|,x_h)=
  \frac{1}{\pi}\sqrt{\frac{C_F}{C_A}}
  \left(|\bm{\kappa}_h|^2\right)^{i\nu-\frac{1}{2}}
  \int_{x_h}^1\frac{\drv x}{x}
  \int_{\frac{x_h}{x}}^1\frac{\drv \eta}{\eta}
  \left(\frac{x\eta}{x_h}\right)^{2i\nu-1}
\end{equation}
  \[ \times \,
  \left[
  \frac{C_A}{C_F}f_g(x)D_g^h\left(\frac{x_h}{x\eta}\right){\cal C}_{gg}
  \left(x,\eta\right)+\sum_{i=q\bar q}f_i(x)D_i^h
  \left(\frac{x_h}{x\eta}
  \right){\cal C}_{qq}\left(x,\eta\right)
  \right.
  \]
  \[ + \,
  \left.D_g^h\left(\frac{x_h}{x\eta}\right)
  \sum_{i=q\bar q}f_i(x){\cal C}_{qg}
  \left(x,\eta\right)+\frac{C_A}{C_F}f_g(x)\sum_{i=q\bar q}D_i^h
  \left(\frac{x_h}{x\eta}\right){\cal C}_{gq}\left(x,\eta\right)
  \right]\, ,
  \]
with the ${\cal C}_{ij}$ partonic coefficients being
\begin{equation}
\stepcounter{appcnt}
\label{Cgg_hadron}
 {\cal C}_{gg}\left(x,\eta\right) =  P_{gg}(\eta)\left(1+\eta^{-2\gamma}\right)
 \ln \left( \frac {|\bm{\kappa}_h|^2 x^2 \eta^2 }{\mu_F^2 x_h^2}\right)
 -\frac{\beta_0}{2}\ln \left( \frac {|\bm{\kappa}_h|^2 x^2 \eta^2 }
 {\mu^2_R x_h^2}\right)
\end{equation}
\[
 + \, \delta(1-\eta)\left[C_A \ln\left(\frac{s_0 \, x^2}{|\bm{\kappa}_h|^2 \,
 x_h^2 }\right) \chi(n,\gamma)
 - C_A\left(\frac{67}{18}-\frac{\pi^2}{2}\right)+\frac{5}{9}n_f
 \right.
\]
\[
 \left.
 +\frac{C_A}{2}\left(\psi^\prime\left(1+\gamma+\frac{n}{2}\right)
 -\psi^\prime\left(\frac{n}{2}-\gamma\right)
 -\chi^2(n,\gamma)\right) \right]
 + \, C_A \left(\frac{1}{\eta}+\frac{1}{(1-\eta)_+}-2+\eta\bar\eta\right)
\]
\[
 \times \, \left(\chi(n,\gamma)(1+\eta^{-2\gamma})-2(1+2\eta^{-2\gamma})\ln\eta
 +\frac{\bar \eta^2}{\eta^2}{\cal I}_2\right)
\]
\[
 + \, 2 \, C_A (1+\eta^{-2\gamma})
 \left(\left(\frac{1}{\eta}-2+\eta\bar\eta\right) \ln\bar\eta
 +\left(\frac{\ln(1-\eta)}{1-\eta}\right)_+\right) \ ,
\]

\begin{equation}
\stepcounter{appcnt}
\label{Cgq_hadron}
 {\cal C}_{gq}\left(x,\eta\right)=P_{qg}(\eta)\left(\frac{C_F}{C_A}+\eta^{-2\gamma}\right)\ln \left( \frac {|\bm{\kappa}_h|^2 x^2 \eta^2 }{\mu_F^2 x_h^2}\right)
\end{equation}
\[
 + \, 2 \, \eta \bar\eta \, T_R \, \left(\frac{C_F}{C_A}+\eta^{-2\gamma}\right)+\, P_{qg}(\eta)\, \left(\frac{C_F}{C_A}\, \chi(n,\gamma)+2 \eta^{-2\gamma}\,\ln\frac{\bar\eta}{\eta} + \frac{\bar \eta}{\eta}{\cal I}_3\right) \ ,
\]

\begin{equation}
\stepcounter{appcnt}
\label{qg}
 {\cal C}_{qg}\left(x,\eta\right) =  P_{gq}(\eta)\left(\frac{C_A}{C_F}+\eta^{-2\gamma}\right)\ln \left( \frac {|\bm{\kappa}_h|^2 x^2 \eta^2 }{\mu_F^2 x_h^2}\right)
\end{equation}
\[
 + \eta\left(C_F\eta^{-2\gamma}+C_A\right) + \, \frac{1+\bar \eta^2}{\eta}\left[C_F\eta^{-2\gamma}(\chi(n,\gamma)-2\ln\eta)+2C_A\ln\frac{\bar \eta}{\eta} + \frac{\bar \eta}{\eta}{\cal I}_1\right] \ ,
\]
and
\begin{equation}
\stepcounter{appcnt}
\label{Cqq_hadron}
 {\cal C}_{qq}\left(x,\eta\right)=P_{qq}(\eta)\left(1+\eta^{-2\gamma}\right)\ln \left( \frac {|\bm{\kappa}_h|^2 x^2 \eta^2 }{\mu_F^2 x_h^2}\right)-\frac{\beta_0}{2}\ln \left( \frac {|\bm{\kappa}_h|^2 x^2 \eta^2 }{\mu^2_R x_h^2}\right)
\end{equation}
\[
 + \, \delta(1-\eta)\left[C_A \ln\left(\frac{s_0 \, x_h^2}{|\bm{\kappa}_h|^2 \, x^2 }\right) \chi(n,\gamma)+ C_A\left(\frac{85}{18}+\frac{\pi^2}{2}\right)-\frac{5}{9}n_f - 8\, C_F \right.
\]
\[
 \left. +\frac{C_A}{2}\left(\psi^\prime\left(1+\gamma+\frac{n}{2}\right)-\psi^\prime\left(\frac{n}{2}-\gamma\right)-\chi^2(n,\gamma)\right) \right] + \, C_F \,\bar \eta\,(1+\eta^{-2\gamma})
\]
\[
 +\left(1+\eta^2\right)\left[C_A (1+\eta^{-2\gamma})\frac{\chi(n,\gamma)}{2(1-\eta )_+}+\left(C_A-2\, C_F(1+\eta^{-2\gamma})\right)\frac{\ln \eta}{1-\eta}\right]
\]
\[
 +\, \left(C_F-\frac{C_A}{2}\right)\left(1+\eta^2\right)\left[2(1+\eta^{-2\gamma})\left(\frac{\ln (1-\eta)}{1-\eta}\right)_+ + \frac{\bar \eta}{\eta^2}{\cal I}_2\right] \; ,
\]

The $s_0$ scale is a BFKL-typical energy-normalization parameter, usually set to $s_0 = \mu_C$.
Furthermore, one has $\bar \eta \equiv 1 - \eta$ and $\gamma \equiv - \frac{1}{2} + i \nu$. The LO DGLAP $P_{i j}(\eta)$ splitting functions read
\begin{eqnarray}
\stepcounter{appcnt}
\label{DGLAP_kernels}
 P_{gq}(z)&=&C_F\frac{1+(1-z)^2}{z} \; , \\ \nonumber
 P_{qg}(z)&=&T_R\left[z^2+(1-z)^2\right]\; , \\ \nonumber
 P_{qq}(z)&=&C_F\left( \frac{1+z^2}{1-z} \right)_+= C_F\left[ \frac{1+z^2}{(1-z)_+} +{3\over 2}\delta(1-z)\right]\; , \\ \nonumber
 P_{gg}(z)&=&2C_A\left[\frac{1}{(1-z)_+} +\frac{1}{z} -2+z(1-z)\right]+\left({11\over 6}C_A-\frac{n_f}{3}\right)\delta(1-z) \; ,
\end{eqnarray}
while the ${\cal I}_{2,1,3}$ functions are
\begin{equation}
\stepcounter{appcnt}
\label{I2}
{\cal I}_2=
\frac{\eta^2}{\bar \eta^2}\left[
\eta\left(\frac{{}_2F_1(1,1+\gamma-\frac{n}{2},2+\gamma-\frac{n}{2},\eta)}
{\frac{n}{2}-\gamma-1}-
\frac{{}_2F_1(1,1+\gamma+\frac{n}{2},2+\gamma+\frac{n}{2},\eta)}{\frac{n}{2}+
\gamma+1}\right)\right.
\end{equation}
\[
 \stepcounter{appcnt}
 \left.
 +\eta^{-2\gamma}\left(\frac{{}_2F_1(1,-\gamma-\frac{n}{2},1-\gamma-\frac{n}{2},\eta)}{\frac{n}{2}+\gamma}-\frac{{}_2F_1(1,-\gamma+\frac{n}{2},1-\gamma+\frac{n}{2},\eta)}{\frac{n}{2} -\gamma}\right)
\right.
\]
\[
 \left.
 +\left(1+\eta^{-2\gamma}\right)\left(\chi(n,\gamma)-2\ln \bar \eta \right)+2\ln{\eta}\right] \; ,
\]
\begin{equation}
\stepcounter{appcnt}
\label{I1}
 {\cal I}_1=\frac{\bar \eta}{2\eta}{\cal I}_2+\frac{\eta}{\bar \eta}\left[\ln \eta+\frac{1-\eta^{-2\gamma}}{2}\left(\chi(n,\gamma)-2\ln \bar \eta\right)\right] \; ,
\end{equation}
and
\begin{equation}
\stepcounter{appcnt}
\label{I3}
 {\cal I}_3=\frac{\bar \eta}{2\eta}{\cal I}_2-\frac{\eta}{\bar \eta}\left[\ln \eta+\frac{1-\eta^{-2\gamma}}{2}\left(\chi(n,\gamma)-2\ln \bar \eta\right)\right] \; .
\end{equation}
Moreover, ${}_2F_1$ stands for the Gauss hypergeometric function.

The \emph{plus~prescription} in Eqs.~(\ref{Cgg_hadron}) and~(\ref{Cqq_hadron}) is given by
\begin{equation}
\label{plus-prescription}
\stepcounter{appcnt}
\int^1_\zeta \drv x \frac{f(x)}{(1-x)_+}
=\int^1_\zeta \drv x \frac{f(x)-f(1)}{(1-x)}
-\int^\zeta_0 \drv x \frac{f(1)}{(1-x)}\; ,
\end{equation}
where $f(x)$ represents a regular-behaved generic function at $x=1$.

\setcounter{appcnt}{0}
\hypertarget{app:NLOJEF}{
\section*{Appendix~B: NLO correction for the light-jet singly off-shell emission function}}
\label{app:NLOJEF}

The analytic expression for the NLO correction to the forward light-jet singly off-shell emission function within the small-cone algorithm reads~\cite{Caporale:2012ih,Colferai:2015zfa}
\begin{equation}
\stepcounter{appcnt}
\label{NLOJEF}
 \hat \F_{J}(n,\nu,|\bm{\kappa}_J|,x_J)=
 \frac{1}{\pi}\sqrt{\frac{C_F}{C_A}}
 \left(|\bm{\kappa}_J|^2 \right)^{i\nu-1/2}
 \int^1_{x_J}\frac{\drv \eta}{\eta}
 \eta^{-\bar\alpha_s(\mu_R)\chi(n,\nu)}
\end{equation}
\[
\times\;
\left\{\sum_{i=q,\bar q} f_i \left(\frac{x_J}{ \eta}\right)\left[\left(P_{qq}(\eta)+\frac{C_A}{C_F}P_{gq}(\eta)\right)
\ln\frac{|\bm{\kappa}_J|^2}{\mu_F^2}\right.\right.
\]
\[
-\;2\eta^{-2\gamma} \ln \frac{{\cal R}}{\max(\eta, \bar \eta)} \,
\left\{P_{qq}(\eta)+P_{gq}(\eta)\right\}-\frac{\beta_0}{2}
\ln\frac{|\bm{\kappa}_J|^2}{\mu_R^2}\delta(1-\eta)
\]
\[
+\;C_A\delta(1-\eta)\left[\chi(n,\gamma)\ln\frac{s_0}{|\bm{\kappa}_J|^2}
+\frac{85}{18}
\right.
\]
\[
\left.
+\;\frac{\pi^2}{2}+\frac{1}{2}\left(\psi^\prime
\left(1+\gamma+\frac{n}{2}\right)
-\psi^\prime\left(\frac{n}{2}-\gamma\right)-\chi^2(n,\gamma)\right)
\right]
\]
\[
+\;(1+\eta^2)\left\{C_A\left[\frac{(1+\eta^{-2\gamma})\,\chi(n,\gamma)}
{2(1-\eta)_+}-\eta^{-2\gamma}\left(\frac{\ln(1-\eta)}{1-\eta}\right)_+
\right]
\right.
\]
\[
\left.
+\;\left(C_F-\frac{C_A}{2}\right)\left[ \frac{\bar \eta}{\eta^2}{\cal I}_2
-\frac{2\ln\eta}{1-\eta}
+2\left(\frac{\ln(1-\eta)}{1-\eta}\right)_+ \right]\right\}
\]
\[
+\;\delta(1-\eta)\left(C_F\left(3\ln 2-\frac{\pi^2}{3}-\frac{9}{2}\right)
-\frac{5n_f}{9}\right)
+C_A\eta+C_F\bar \eta
\]
\[
\left.
+\;\frac{1+\bar \eta^2}{\eta}
\left(C_A\frac{\bar \eta}{\eta}{\cal I}_1+2C_A\ln\frac{\bar\eta}{\eta}
+C_F\eta^{-2\gamma}(\chi(n,\gamma)-2\ln \bar \eta)\right)\right]
\]
\[
+\;f_{g}\left(\frac{x_J}{\eta}\right)\frac{C_A}{C_F}
\left[
\left(P_{gg}(\eta)+2 \,n_f \frac{C_F}{C_A}P_{qg}(\eta)\right)
\ln\frac{|\bm{\kappa}_J|^2}{\mu_F^2}
\right.
\]
\[
\left.
-\;2\eta^{-2\gamma} \ln \frac{{\cal R}}{\max(\eta, \bar \eta)} \left(P_{gg}(\eta)+2 \,n_f P_{qg}(\eta)\right)
-\frac{\beta_0}{2}\ln\frac{|\bm{\kappa}_J|^2}{4\mu_R^2}\delta(1-\eta)
\right.
\]
\[
\left.
+\; C_A\delta(1-\eta)
\left(
\chi(n,\gamma)\ln\frac{s_0}{|\bm{\kappa}_J|^2}+\frac{1}{12}+\frac{\pi^2}{6}
\right.\right.
\]
\[
\left.
+\;\frac{1}{2}\left[\psi^\prime\left(1+\gamma+\frac{n}{2}\right)
-\psi^\prime\left(\frac{n}{2}-\gamma\right)-\chi^2(n,\gamma)\right]
\right)
\]
\[
+\,2C_A (1-\eta^{-2\gamma})\left(\left(\frac{1}{\eta}-2
+\eta\bar\eta\right)\ln \bar \eta + \frac{\ln (1-\eta)}{1-\eta}\right)
\]
\[
+\,C_A\, \left[\frac{1}{\eta}+\frac{1}{(1- \eta)_+}-2+\eta\bar\eta\right]
\left((1+\eta^{-2\gamma})\chi(n,\gamma)-2\ln\eta+\frac{\bar \eta^2}
{\eta^2}{\cal I}_2\right)
\]
\[
\left.\left.
+\,n_f\left[\, 2\eta\bar \eta \, \frac{C_F}{C_A} +(\eta^2+\bar \eta^2)
\left(\frac{C_F}{C_A}\chi(n,\gamma)+\frac{\bar \eta}{\eta}{\cal I}_3\right)
-\frac{1}{12}\delta(1-\eta)\right]\right]\right\} \; ,
\]
with ${\cal R}$ denoting the jet-cone radius.

\end{appendices}

\bibliographystyle{elsarticle-num}

\bibliography{references}

\end{document}